\documentclass{aa}  

\usepackage{bm}
\usepackage{siunitx}
\DeclareSIUnit\erg{erg}
\DeclareSIUnit\Mpc{Mpc}

\usepackage{multirow}
\usepackage{subcaption}
\usepackage{tabularx}
\newcolumntype{C}{>{\centering\arraybackslash}X}

\usepackage{graphicx}
\usepackage[dvipsnames]{xcolor}
\usepackage{enumitem} 

\usepackage{txfonts}

\usepackage{color}
\usepackage{ulem}
\usepackage{natbib, twoopt}
\usepackage[hyphenbreaks]{breakurl}
\usepackage[breaklinks]{hyperref}
\bibpunct{(}{)}{;}{a}{}{,}
\hypersetup{
  colorlinks,
  citecolor=NavyBlue,
  linkcolor=NavyBlue,
  urlcolor=NavyBlue,
}
\makeatletter
  \newcommandtwoopt{\citeads}[3][][]{\href{http://adsabs.harvard.edu/abs/#3}%
    {\def\hyper@linkstart##1##2{}%
     \let\hyper@linkend\@empty\citealp[#1][#2]{#3}}}
  \newcommandtwoopt{\citepads}[3][][]{\href{http://adsabs.harvard.edu/abs/#3}%
    {\def\hyper@linkstart##1##2{}%
     \let\hyper@linkend\@empty\citep[#1][#2]{#3}}}
  \newcommandtwoopt{\citetads}[3][][]{\href{http://adsabs.harvard.edu/abs/#3}%
    {\def\hyper@linkstart##1##2{}%
     \let\hyper@linkend\@empty\citet[#1][#2]{#3}}}
  \newcommandtwoopt{\citeyearads}[3][][]%
    {\href{http://adsabs.harvard.edu/abs/#3}
    {\def\hyper@linkstart##1##2{}%
     \let\hyper@linkend\@empty\citeyear[#1][#2]{#3}}}
\makeatother

\def\ngal{116}
\def\ngalfit{95}
\def\OII{\hbox{[{\rm O}{\sc \,ii}]}}
\def\galpak{\texttt{GalPaK$^\texttt{3D}$}}
\def\LCDM{$\Lambda\mbox{CDM}$}

\begin{document}
\defcitealias{Richard_21}{R21}
\defcitealias{ManceraPina_26}{MP26}

   \title{MUSE-DARK-II: 3D morpho-kinematic modelling of lensed galaxies}

   \subtitle{Tully-Fisher relation of $z \sim 1$ star-forming galaxies}

   \author{Alexandre Jeanneau \inst{1}
          \and
          Johan Richard \inst{1}
          \and
          Nicolas  F. Bouché \inst{1}
          \and
          Davor Krajnović \inst{2}
          \and
          Bianca-Iulia Ciocan \inst{1}
          \and
          Jonathan Freundlich \inst{3}
          \and
          Benoît Epinat \inst{4, 5}
          \and
          Thierry Contini \inst{6}}

   \institute{Universite Claude Bernard Lyon 1, CRAL UMR5574, ENS de Lyon, CNRS, Villeurbanne, F-69622, France\\
              \email{alexandre.jeanneau@univ-lyon1.fr}
              \and
              Leibniz-Institut für Astrophysik Potsdam (AIP), An der Sternwarte 16, 14482 Potsdam, Germany
              \and
              Observatoire Astronomique de Strasbourg, Université de Strasbourg, CNRS, UMR 7550, 67000 Strasbourg, France
              \and
              French-Chilean Laboratory for Astronomy, IRL 3386, CNRS and Universidad de Concepción, Departamento de Astronomía, Barrio Universitario s/n, Concepción, Chile
              \and
              Aix Marseille Univ, CNRS, CNES, LAM, Marseille, France
              \and
              IRAP, Université de Toulouse, CNRS, CNES, 14 Av. Edouard Belin, 31400 Toulouse, France
              }

   \date{Received XX XX, 2025}

 
\abstract{Extending local kinematic studies to earlier cosmic times is valuable to understand how galaxies evolve in relation to their dark matter haloes. In a series of papers on lensed kinematics, we seek to combine the sensitivity of 3D forward modelling to low signal-to-noise ratio outskirts with the enhanced spatial resolution provided by cluster lensing. In this first paper, we (i) present and validate our methodology, which directly constrains the source parameters by incorporating lensing deflections into the \galpak{} forward-modelling algorithm, and (ii) investigate the evolution of the stellar-mass and baryonic-mass Tully–Fisher relations (sTFR and bTFR) since $z \sim 1$ as a demonstration. We define a robust sample of strongly lensed star-forming galaxies (SFGs) from the MUSE Lensing Cluster survey, spanning magnifications $\mu = 1.4 - 12.4$ and stellar masses $M_\star = 10^{8.1} - 10^{10.3} M_\odot$. Using a series of mock galaxies representative of our sample, we find that our method is significantly more reliable at recovering morpho-kinematic properties than approaches that ignore differential magnification, even for relatively modest magnifications ($\mu < 6$). Restricting the analysis to \ngalfit{} rotationally supported SFGs with well-constrained velocities, we find a significant evolution of the sTFR zero-point ($\Delta b^\mathrm{sTFR} = -0.42^{+0.05}_{-0.05}~\mathrm{dex}$ in stellar mass) but no detectable evolution of the bTFR zero-point ($\Delta b^\mathrm{bTFR} = 0.00^{+0.06}_{-0.06}~\mathrm{dex}$ in baryonic mass) relative to $z \approx 0$. Our results are consistent with a mild evolution of the stellar-to-halo mass ratio and support the view that the sTFR has evolved only weakly over the past $\sim 8$ Gyr, aside from shifts driven by the redshift dependence of halo-defining quantities such as the critical density and overdensity. The absence of detectable evolution in the bTFR zero-point suggests that the increasing contribution of cold gas mass at higher redshift fully compensates the evolution observed in the stellar component alone.}


   \maketitle
%
\section{Introduction}
    Rotation curves (RCs) provide one of the most direct probes of the total mass profile of a galaxy, which can be contrasted with the visible distribution of its stars and gas. Decades of kinematic studies of nearby disc galaxies, enabled by high-quality RCs, have established a set of empirical results that remain central to current discussions on the physics underlying galaxy evolution. The discovery that the RCs of nearby disc galaxies remain nearly flat out to tens of kiloparsecs \citep{Bosma_78, Rubin_78}, far beyond the radius where a Keplerian decline is expected from visible matter alone, has been key in advancing the standard \LCDM{} cosmology \citep[e.g.][]{Mo_98} as well as alternative gravity theories deviating from Newtonian dynamics at low accelerations \citep[e.g.][]{Sanders_02, Famaey_12}. Another remarkable finding is their apparent regularity, which manifests in the capacity to generally represent RCs using a universal function of the luminosity and optical scale length \citep[e.g.][]{Persic_96} as well as in their adherence to tight scaling relations. These include global relations such as the Tully–Fisher \citep[TFR; e.g.][]{Tully_77, Bell_01, McGaugh_00} and Fall \citep{Fall_80} relations which extend over several orders of magnitude in stellar mass, as well as local relations such as the radial acceleration relation \citep[e.g.][]{McGaugh_16}. On small scales, the slowly rising RCs of low-mass, dark matter (DM) dominated galaxies imply shallow central DM densities and slopes, challenging the expectations from DM-only halo profiles \citep[][for a review]{Bullock_17}.
    
    Extending local kinematic studies to earlier cosmic times is valuable to understand how galaxies evolve in relation to their DM haloes. Over the past two decades, this effort has been facilitated by sensitive 3D spectrographs across the optical (e.g. VLT/MUSE, Keck/KCWI), near-infrared (e.g. VLT/SINFONI, VLT/KMOS, Keck/OSIRIS, JWST/NIRSpec) and millimeter (e.g. ALMA, NOEMA) domains. Observations now reveal that disc galaxies at intermediate redshifts ($0.5 < z < 3$) are reasonably regular albeit dynamically hotter \citep[e.g.][]{Schreiber_06} and thicker, with scaling relations such as the TFR \citep[][and references therein]{Turner_17, Sharma_24} or Fall \citep[e.g.][]{Burkert_16, Harrison_17, Swinbank_17, Bouche_21} relations already established \citep[][for a review]{Schreiber_20}. However, there is no consensus yet as to the outer shape of intermediate-redshift RCs, which is key to constrain DM halo masses: some studies report flat or rising profiles \citep[e.g.][]{Sharma_21, Puglisi_23}, while others also find declining profiles at large radii \citep[e.g.][]{Lang_17, Genzel_17, Genzel_20, NestorShachar_23}. The evolution of the stellar-mass TFR is similarly unsettled. Recently, \citet{ManceraPina_26} reported moderate evidence for a shallower slope at $z \sim 1$ compared to the $z \approx 0$ relation of \citet{Marasco_25}, in tension with expectations if the two galaxy populations share a simple progenitor–descendant relation. A major limitation is that most existing samples span a relatively narrow stellar-mass range ($M_\star \gtrsim 10^{9.5-10} M_\odot$; for extensions to lower masses see \citealt{Contini_16, AbrilMelgarejo_21, Mercier_22}), limiting constraints on possible slope evolution. Even when the slope is fixed to local values, conclusions on the evolution of the zero-point range from a mild evolution if any \citep[e.g.][]{Pelliccia_17, Tiley_19}, to a moderate evolution \citep[e.g.][]{Ubler_17}. However, \citet{Turner_17} pointed out that many of these studies can be reconciled with a moderate evolution of the TFR (corresponding to $\simeq -0.4~\mathrm{dex}$ in stellar mass at $z \sim 1$, relative to $z \approx 0$), once the contribution of pressure support to the dynamics of distant star-forming galaxies (SFGs) is properly included. Finally, constraints on DM density profiles at these epochs remain scarce. Resolving the inner RC calls for high spatial resolution, while constraining the outer RC requires deep observations, making such measurements observationally demanding. With the exception of \citet{Bouche_22}, \citet{Ciocan_26} and \citet[][which relies on slitless spectroscopy]{Danhaive_26}, most existing studies focus on relatively massive SFGs with $M_\star \gtrsim 10^{10} M_\odot$ \citep[e.g.][]{Genzel_20, NestorShachar_23}.
    
    Disentangling physical evolution from observational biases remains challenging. Kinematic measurements rely primarily on molecular or warm ionised gas, sometimes supplemented by stellar kinematics \citep{Guerou_17, Straatman_22, Ubler_24, MunozLopez_24}. Unlike HI, those tracers generally do not reach far beyond the optical disc, but studies of individual galaxies in HI are not yet feasible because of the intrinsic faintness of the $21~\mathrm{cm}$ line. In addition, distant SFGs are apparently and intrinsically fainter, smaller \citep[e.g.][]{Wel_14, Nedkova_21} and more turbulent \citep[e.g.][]{Wisnioski_25}. Marginal spatial resolution leads to beam smearing, which smooths out velocity gradients and inflates the observed velocity dispersion \citep[][for an extensive analysis]{Epinat_08}, while cosmological dimming limits the signal-to-noise ratio (S/N) achievable in typical exposure times. Forward-modelling tools that include observational effects, such as \galpak{} \citep{Bouche_15a, Bouche_15b}, \texttt{DysmalPy} \citep[][and references therein]{Price_21, Lee_25}, \texttt{$^{3\mathrm{D}}$Barolo} \citep{DiTeodoro_15}, or \texttt{qubefit} \citep{Neeleman_20, Neeleman_21}, help mitigate those limitations by leveraging the full 3D information in the data \citep[see][for comparisons]{Lee_25, Yttergren_25}. Nonetheless, limited spatial resolution can still wash out kinematic features \citep[see the validation tests in][]{Bouche_15b, DiTeodoro_15} and sometimes result in misclassifications \citep{Leethochawalit_16, Simons_19}. Combined with typically small sample sizes, this complicates comparisons with both local and other high-redshift datasets.
    
    Attempts to overcome these limitations have used either adaptive optics \citep[e.g.][]{Schreiber_18} or gravitational lensing \citep[e.g.][]{Girard_20} to achieve $\simeq 1-4~\mathrm{kpc}$ resolution at high redshift. Near-diffraction-limited adaptive optics observations remain restricted to small samples and/or relatively massive SFGs, since deep integrations are performed on a target-by-target basis. Cluster-lensed observations do not face this constraint, but they require additional modelling to recover intrinsic source properties. Previous kinematic studies of lensed SFGs have mostly analysed source-plane reconstructions of the datacubes \citep[e.g.][]{Jones_10, Jones_13, Swinbank_11, Livermore_15}, or fitted kinematic parameters directly in the image plane \citep[e.g.][]{DiTeodoro_18, Girard_18, Girard_20}, and in some cases included a global correction to account for the lensed axis ratio \citep[e.g.][]{Mason_17}. A handful of studies have incorporated lensing deflections directly in their kinematic modelling. \citet{Patricio_18} proposed a forward-modelling approach accounting for the point-spread function (PSF) and lensing deflections on resolved velocity and metallicity maps. \citet{Rizzo_18} developed an algorithm that jointly infers the lens mass model and the source kinematics from the same datacube, although this was applied to galaxy or group-scale lenses with comparatively simple lens mass distributions. Another tool that includes the modelling of galaxy-scale lenses as well as source kinematics is described in \citet{Chirivi_20}. Finally, we highlight \citet{Liu_23} who developed a lensing transformation module to complement the \texttt{DysmalPy} 3D algorithm.
    
    In this series of papers on lensed kinematics, we combine the capacity of 3D algorithms to collectively exploit peripheral (hence, low S/N) spaxels with the enhanced spatial resolution enabled by cluster gravitational lensing to study the kinematics of intermediate-redshift galaxies. We extend \galpak{} to include lensing deflections directly in the forward model, enabling direct comparison between the model and the observed data, rather than intermediate products (reconstructed datacubes or moment maps) whose correlated pixels are known to introduce systematics\footnote{As an example, \citet{Posti_22} demonstrates that treating (blank field) velocity measurements along a RC as independent leads to biased parameter estimates and underestimated errors. Building on this approach, \citet{Chase_25} show that accounting for correlations between RC datapoints is critical to determine whether RCs from the SPARC database \citep{Lelli_16b} statistically favour cored or cuspy DM haloes.}. We model the kinematics of a sample of \ngal{} \OII{} emitters drawn from the MUSE Atlas of Lensing Clusters \citep[][hereafter R21]{Richard_21}, spanning the redshift range $0.5 < z < 1.5$. This first paper focuses on the methodology and sample definition, and investigates the evolution of the stellar-mass and baryonic-mass Tully–Fisher relations (sTFR and bTFR) since $z \sim 1$ as a demonstration.
    
    This paper is organized as follows. We begin with a brief overview of the MUSE Lensing Cluster Survey and the Frontier Fields ancillary data (Sect.~\ref{sect:data}). We then describe the implementation of our strong-lensing kinematic modelling with \galpak{} (Sect.~\ref{sect:galpak_SLextension}) and assess its recovery performance using mock datacubes (Sect.~\ref{sect:recovery_perf}). Next, we define and characterise our kinematic sample (Sect.~\ref{sect:sample}) and use it to investigate the sTFR and bTFR at intermediate redshift (Sect.~\ref{sect:results}). The results are discussed in Sect.~\ref{sect:discussion}, and our conclusions are summarised in Sect.~\ref{sect:conclusions}.
    
    Throughout this analysis, we assume a standard \LCDM{} cosmology with $\Omega_{\rm m} = 0.3$, $\Omega_{\Lambda} = 0.7$, and $H_0 = \qty{70}{\km.\s^{-1}.\Mpc^{-1}}$. Unless specified otherwise, we adopt a Bayesian approach and report best-fit parameters as the median of their posterior distribution. Stellar masses and star formation rates (SFRs) are derived assuming a \citet{Chabrier_03} initial mass function (IMF), or converted to this IMF where necessary.

\section{Data and cluster mass models}
\label{sect:data}
    The sample of \OII{} emitters used in this analysis is a subsample of the MUSE Atlas of Lensing Clusters \footnote{\url{https://cral-perso.univ-lyon1.fr/labo/perso/johan.richard/MUSE_data_release/}} \citepalias{Richard_21}, a legacy program targeting the central region of 19 massive galaxy clusters with the MUSE instrument \citep{Bacon_10}. In this study, we focus on four clusters that are part of the Frontier Fields, which benefit from extensive ancillary data: Abell 2744, Abell 370, Abell S1063, and MACS0416. We summarise the key features of the MUSE observations and ancillary datasets below and in Tab.~\ref{tab:clusters}.

        \subsection{The MUSE parent sample of Lensing Clusters }
            The MUSE Atlas of Lensing Clusters combines data acquired with integral-field spectroscopy (IFS) spanning the $475 - \qty{930}{\nano\metre}$ range with a spatial sampling of $0.2'' \times 0.2''$, a spectral sampling of $1.25~\AA{}$, and a spectral resolution from $\mathcal{R} = 1800$ to $\mathcal{R} = 3600$. The final datacubes combine exposures taken in both standard and ground-layer adaptive optics \citep{Strobele_12} modes, yielding effective integration times of $2-15~\mathrm{hr}$ per pointing. For the four Frontier Fields clusters, the resulting PSF has a median full width at half maximum (FWHM) of $0.65''$ at $7000~\AA{}$, corresponding to an intrinsic spatial resolution of $5.2~\mathrm{kpc}/\sqrt{\mu}$ and a sampling of $1.6~\mathrm{kpc}/\sqrt{\mu}$ at $z \sim 1$, where $\mu$ is the lensing magnification factor. The multiple-image regions were covered with four adjacent pointings for Abell 370 and Abell 2744, and two pointings for Abell S1063 and MACS0416. Data reduction largely followed the standard procedure of \citet{Weilbacher_20}, with some adaptations to cope with the high density of extended bright sources in lensing cluster fields. The observing and data reduction strategies are detailed in \citetalias{Richard_21} for the first twelve clusters, including Abell 2744, Abell 370, and MACS0416, and in \citet{Claeyssens_22} for five additional clusters, including Abell S1063.

        \subsection{Ancillary data}
            In addition to the MUSE data products, this study relies on the wealth of observations targetting the Frontier Fields. The HFF-DeepSpace project \citep{Shipley_18} combined Hubble Space Telescope (HST) ACS and WFC3 imaging with $K_s$-band observations from the KIFF survey \citep{Brammer_16}, obtained using VLT/HAWK-I and Keck/MOSFIRE. Together, those datasets provide uniform multi-wavelength coverage in up to 17 filters, spanning the UV to the NIR, and are supplemented by post-cryogenic Spitzer/IRAC imaging at \qty{3.6}{\micro\metre} and \qty{4.5}{\micro\metre} (and archival \qty{5.8}{\micro\metre} and \qty{8.0}{\micro\metre} data where available). The catalogues also include images where the light of bright cluster galaxies (BCGs) has been modelled and subtracted, enabling the detection and analysis of faint background sources. For the purposes of this work, we make use of the HFF-DeepSpace photometric catalogues along with BCG-subtracted images sampled at $0.06''$ per pixel and their corresponding empirical PSFs obtained by stacking isolated and unsaturated stars.

        \subsection{Cluster mass models}
            To study the intrinsic properties of the lensed galaxies, we use the parametric cluster mass models from \citet{Mahler_18}, \citet{Lagattuta_19},  \citetalias{Richard_21} and \citet{Beauchesne_24} (see Tab.~\ref{tab:clusters}), constructed with the public \texttt{Lenstool} software \citep{Kneib_96, Jullo_07}. These models are constrained by numerous multiple-image systems identified in the MUSE and HST catalogs, including both spectroscopically confirmed sources and continuum sources in unambiguous lensing configurations.
            
            Each cluster mass model is built iteratively using pseudo-isothermal elliptical potentials, starting from a few cluster-scale potentials optimised to reproduce the most secure multiple-image systems, and progressing to galaxy-scale potentials whose shapes are scaled according to the light distribution of cluster members (assuming a constant mass-to-light ratio). For the four Frontier Fields clusters, the models are constrained with $29 - 71$ secure multiple-image systems, achieving a typical error of $0.6 - 0.8''$ rms in reproducing the observed images. Statistical uncertainties on the model parameters are estimated via Markov Chain Monte Carlo (MCMC) sampling of the posterior distribution. From these models, we derive two key products for our analysis: lensing magnification maps and deflection maps, which provide the angular displacement between image plane and source plane positions at a given source redshift, allowing us to correct both flux magnification and shape distortions in the lensed galaxies.

\section{\galpak{} Strong Lensing Extension}
\label{sect:galpak_SLextension}

    \begin{figure}[]
        \centering
        \includegraphics[width=0.8\hsize]{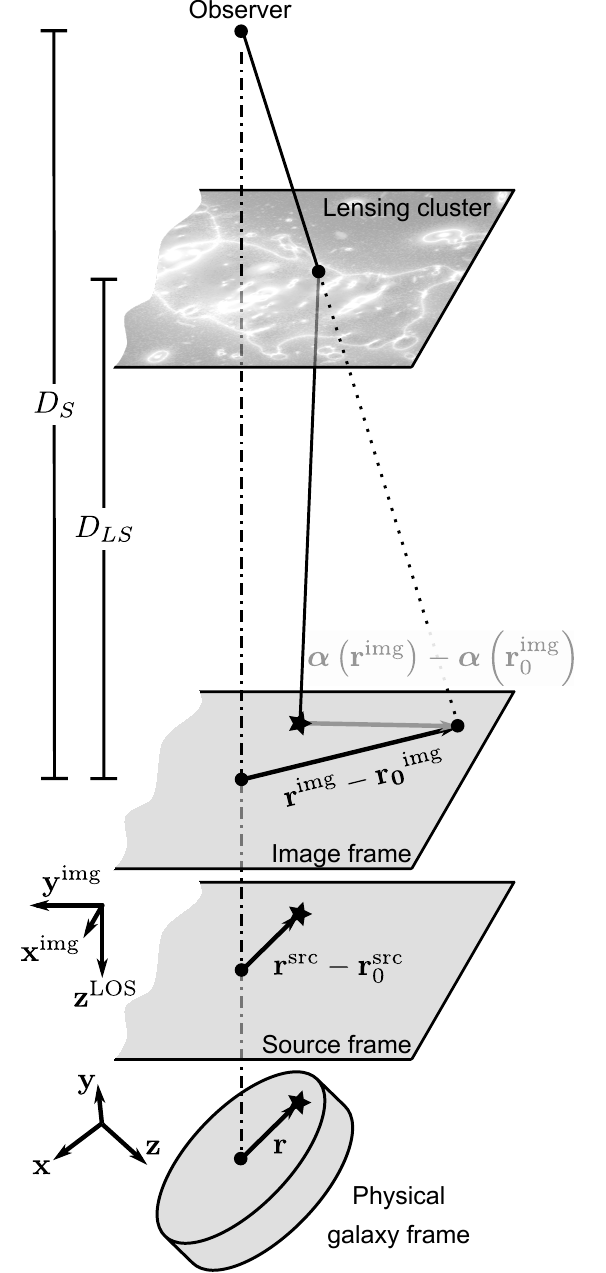}
        \caption{Schematic view of the lensing configuration and coordinate frames. The observed image-frame coordinates $\mathbf{r}^\mathrm{img} - \mathbf{r}_0^\mathrm{img}$ are mapped to source-frame coordinates $\mathbf{r}^\mathrm{src} - \mathbf{r}_0^\mathrm{src}$ via the lens equation $\mathbf{r}^\mathrm{src} = \mathbf{r}^\mathrm{img} - \bm{\alpha}(\mathbf{r}^\mathrm{img})$. The source-frame coordinates correspond to positions $\mathbf{r}$ in the physical galaxy frame, where parametric flux and velocity profiles are defined. These coordinate transformations do not alter the model grid, which remains aligned and regularly sampled along the IFS spatial axes and the LOS. This schematic is inspired by \citet{Bartelmann_01} and \citet{DiTeodoro_15}.}
        \label{fig:coordinate_transforms}
    \end{figure}
    
    In this paper, we extract the intrinsic morphology and kinematics of distant SFGs using an extension of the 3D forward-modelling tool \galpak{}, which simulates lensed emission-line datacubes from a parametric rotating disc model. Building on the standard \galpak{} algorithm, we adapt the forward-modelling coordinate transformations to include lensing deflections. The extension proceeds as follows (Fig.~\ref{fig:coordinate_transforms}).

    \paragraph{Image-frame coordinate cube construction} First, we generate a 3D coordinate cube in the image frame $\mathbf{R}^\mathrm{img} = (\mathbf{r}^\mathrm{img}, z^\mathrm{LOS} )$, where $\mathbf{r}^\mathrm{img} = (x^\mathrm{img}, y^\mathrm{img})$ are the sky coordinates, defined on an oversampling of the observed IFS datacube grid, and $z^\mathrm{LOS}$ is the spatial coordinate along the line of sight (LOS). The model grid is typically oversampled by a factor of two to mitigate discretisation errors. We emphasise that subsequent transformations act on the spatial coordinates stored in the cube, while the grid itself remains aligned and regularly sampled along the IFS spatial axes and the LOS.
    
    \paragraph{Image-plane centering} Coordinates are then expressed relative to the galaxy centre, $\mathbf{R}_0^\mathrm{img} = (\mathbf{r_0}^\mathrm{img}, z_0^\mathrm{LOS})$. Here, $\mathbf{r}_0^\mathrm{img} = (x_0^\mathrm{img}, y_0^\mathrm{img})$ corresponds to the sky coordinates of the apparent galaxy centre fitted by \galpak{}, and $z_0^\mathrm{LOS}$ is set such that the galaxy centre lies in the middle of the model cube along the LOS. For multiply imaged systems, fitting the apparent centre of one counter-image is sufficient to reproduce the full system within the accuracy of the lens model.
    
    \paragraph{Source-plane mapping} Using the lens equation, the coordinates stored in $\mathbf{R}^\mathrm{img} - \mathbf{R}_0^\mathrm{img}$ are mapped to the source plane and expressed relative to the galaxy centre as $\mathbf{R}^\mathrm{src} - \mathbf{R}_0^\mathrm{src} = (\mathbf{r}^\mathrm{src} - \mathbf{r}_0^\mathrm{src}, z^\mathrm{LOS} - z_0^\mathrm{LOS})$:
    
    \begin{equation}
        \mathbf{r}^\mathrm{src} - \mathbf{r}_0^\mathrm{src} = \left[ \mathbf{r}^\mathrm{img} - \mathbf{r_0}^\mathrm{img} \right] - \left[\bm{\alpha} \left( \mathbf{r}^\mathrm{img} \right)  - \bm{\alpha} \left( \mathbf{r}_0^\mathrm{img} \right) \right] ,
    \end{equation}
    
    where $\bm{\alpha}$ is the user-provided deflection angle resulting from the cluster lens model. Lens models commonly include reduced deflection maps $\bm{\widehat{\alpha}}$, defined such that $\bm{\alpha} = (D_{LS} / D_{S}) \ \bm{\widehat{\alpha}}$, where $D_{LS}$ and $D_{S}$ are the angular-diameter distances from the lens to the source and from the observer to the source, respectively. We use a bilinear interpolation to derive $\bm{\alpha} (\mathbf{r}_0^\mathrm{img})$ from the pixelated deflection maps. The result is a cube storing source-frame coordinates, regularly sampled in the image frame.

    \paragraph{Rotation into the physical galaxy frame} The source-frame coordinates are then rotated into the physical galaxy frame, $\mathbf{R} = \left(\mathbf{r}, z \right)$, where $\mathbf{r} = \left(x, y\right)$ lies in the galaxy midplane and $z$ is the coordinate along the direction perpendicular to it. This transformation is performed by applying two successive 3D rotations to the source-frame coordinates: a rotation $\mathcal{R}_z(\mathrm{PA})$ about the $z$-axis to account for the position angle (PA), followed by a rotation $\mathcal{R}_x(i)$ about the $x$-axis to account for the inclination $i$:
    
    \begin{equation}
        \mathbf{R} = \mathcal{R}_x\left( i \right) \, \mathcal{R}_z\left( \mathrm{PA} \right) \, \left(  \mathbf{R}^\mathrm{src} - \mathbf{R}_0^\mathrm{src}  \right).
    \end{equation}

    The resulting coordinates are expressed in the physical galaxy frame. The following steps then proceed as in the standard \galpak{} algorithm, except that we continue to use an oversampled model cube to minimise discretisation errors. We summarise them for self-consistency.

    \paragraph{Flux and kinematic profile assignment} Parametric flux and velocity profiles are applied to the galaxy-frame coordinates cube, that is, each pixel of the cube is allocated a flux or a rotational velocity vector corresponding to its coordinates in the galaxy frame. A flux cube is created from a radial profile $I(\mathbf{r})$ and a vertical profile $I(z)$ relying on a half width at half maximum $h_z$. The rotational velocity field $\mathbf{V}_\perp(\mathbf{R}) = (v_x, v_y, 0)$ is generated from a rotation velocity profile $v_\perp\left(\mathbf{r}, z\right)$ and assumes circular orbits. The RC can be modelled using a simple parametrisation, such as the arctan model \citep{Courteau_97} adopted in this paper, or a more detailed disc-halo decomposition \citep{Bouche_22, Ciocan_26}.
    
    \paragraph{Derivation of moment maps} The flux cube and velocity field are used to generate flux, LOS velocity and LOS velocity dispersion moment maps.
    \begin{itemize}
        \item The flux map is obtained directly by collapsing the flux cube along the LOS.
        \item The LOS velocity map is derived from a LOS velocity cube. At each position, the LOS component of the velocity field is computed as:

        \begin{equation}
            v^\mathrm{LOS}\left(\mathbf{R}^\mathrm{img} - \mathbf{R}_0^\mathrm{img}\right) = v_\perp\left(\mathbf{r}, z\right) \left( \frac{x}{r} \right) \sin i + z_0^\lambda \Delta v,
        \end{equation}

        following the coordinate transformations of the previous steps. $z_0^\lambda$ is the systemic spectral position (fitted by \galpak{}), and $\Delta v$ is the velocity increment per spectral pixel. The LOS velocity map is then obtained as the flux-weighted mean of this $v^\mathrm{LOS}$ cube along the LOS.

        \item The LOS velocity dispersion map combines three contributions in quadrature: from the disc self-gravity $\sigma_d(r)$ (with $\sigma_d(r)/h_z = V_\perp(r)/r$ for thick discs); a projection term due to the blending of different velocities along the LOS, computed from the flux-weighted variance of the $v^\mathrm{LOS}$ cube along the LOS; and a constant and isotropic floor $\sigma_0$ fitted by \galpak{}.
    
    \end{itemize}
    
    \paragraph{Model cube construction} A model cube in observational coordinates $I_\mathrm{mod} (\mathbf{r}^\mathrm{img}, z^\lambda)$ is built by assigning, at each spaxel, a Gaussian line (or doublet) profile with the flux, velocity, and velocity dispersion defined by the corresponding moment maps. This model cube is free from instrumental resolution.
    
    \paragraph{Convolution with the instrumental response and comparison to data} Finally, the model cube is convolved with the instrumental response: a 3D kernel combining the PSF and the line spread function (LSF). The oversampled cube is re-binned before being compared with the data $I_\mathrm{obs} (\mathbf{r}^\mathrm{img}, z^\lambda )$.

    With this end-to-end modelling approach, the morphological and kinematic parameters inferred by \galpak{} are implicitly corrected for lensing distortions and observational effects such as beam smearing.

    Among the algorithms implemented in \galpak{}, we adopt the \texttt{MultiNest} \citep{Feroz_09, Buchner_16} Bayesian inference tool to sample the posterior distribution of the parameters $\bm{\theta}$ given the data $\mathcal{D}$. This sampler provides an efficient and robust sampling strategy for high-dimensional parameter spaces. Throughout this paper, the runs are performed with 200 live points, a sampling efficiency of 0.8, and an evidence tolerance of 0.5. We define a likelihood function as $\mathcal{L} = \mathcal{P} \left(\mathcal{D} | \bm{\theta} \right) = \exp(- \frac{1}{2} \chi_\mathrm{3D}^2)$, with:
    
    \begin{equation}
        \chi_\mathrm{3D}^2 = \sum_{\mathbf{r}^\mathrm{img}, z^\lambda} \frac{\left[ I_\mathrm{obs} \left(\mathbf{r}^\mathrm{img}, z^\lambda \right) - I_\mathrm{mod} \left(\mathbf{r}^\mathrm{img}, z^\lambda \right) \right]^2}{\mathrm{Var}\left[ I_\mathrm{obs} \right]\left(\mathbf{r}^\mathrm{img}, z^\lambda \right)}. 
    \end{equation}
    
    The posterior distribution is then given by the Bayes theorem $\mathcal{P} \left(\bm{\theta} | \mathcal{D}\right) \propto \mathcal{P} \left(\mathcal{D} | \bm{\theta} \right) \mathcal{P} \left( \bm{\theta} \right)$, where $\mathcal{P} \left( \bm{\theta} \right)$ is the overall prior corresponding to the product of the individual priors.

\section{Recovery performance using mock datacubes}
\label{sect:recovery_perf}
    In this section, we examine:
    \begin{enumerate}[label=(\roman*)]
        \item the robustness of our method when applied to datacubes with limited S/N, spatial and spectral resolution;
        \item how our results compare with those obtained using a kinematic model that does not account for differential magnification, in other words, when parameters are inferred in the image frame and, where relevant, simply rescaled by the magnification factor \citep[e.g.][]{DiTeodoro_18, Girard_18, Girard_20}.
    \end{enumerate}

    To this end, we evaluate the performance of the \galpak{} Strong Lensing Extension on a series of mock datacubes representative of our sample (Sect.~\ref{sect:sample}), generated with its forward-modelling capability.

    \subsection{Datacube generation}
        Synthetic galaxies are assigned a redshift drawn from a uniform distribution and stellar masses sampled from the star-forming stellar mass function at $z \sim 1$ \citep{Weaver_23}. Other physical properties (e.g. effective radius, SFR, velocity, velocity dispersion) are set by scaling relations and their scatter, following a procedure comparable with \cite{Lee_25}.
    
        The \OII{} emission, used as the kinematic tracer, is assumed to follow the stellar disc. We adopt an exponential surface brightness profile (Sérsic index $n = 1$) with an effective radius $R_e$ from the stellar mass–size relation of \citet{Nedkova_21}, corrected to rest-frame I band \citep{Wel_14}, and a Gaussian vertical profile with scale height $h_z/\sqrt{2\ln(2)} = 0.13 R_e$ \citep{Lian_24, Tsukui_25}.
        
        The RC is described by an arctan function \citep{Courteau_97}:
        \begin{equation}
        v_c(r) = \frac{2}{\pi} v_\mathrm{max} \arctan\left(\frac{r}{r_t}\right),
        \end{equation}
        where $v_\mathrm{max}$ is the asymptotic velocity and $r_t$ the turnover radius. We report velocities at $1.8 R_e$ (the radius enclosing 80\% of the total light for an exponential disc). In practice, we draw $v_{1.8} = v_c(1.8 R_e)$ from the local TFR reported by \citet{Turner_17} and based on the reference sample of \citet{Reyes_11}, and use it to derive $v_\mathrm{max}$. The turnover radius is set to $r_t = R_d/1.8$, with $R_d = R_e/1.68$ the disc scale length, as derived by \citet{Bouche_15b} from the observed scaling between galaxy size and inner velocity gradient \citep{Amorisco_10}.

        The typical velocity dispersions of $z \sim 1$ galaxies ($10-60~\mathrm{km/s}$, \citealt{Wisnioski_25}) indicate that turbulent motions provide significant pressure support to the ionised gas, counteracting gravity. As a result, the observed rotation velocity traced by \OII{} is lower than the circular velocity, that is, the velocity a test particle would have if it were supported purely by gravity. We therefore include a pressure-support correction to the circular velocity profile, following the prescription of \citet{Dalcanton_10} for an exponential radial gas distribution:
        
        \begin{equation}
        v_\perp(r)^2 = v_c(r)^2 - 0.92 \sigma_r^2(r) \left( \frac{r}{R_d} \right), 
        \end{equation}
        where $\sigma_r^2(r) \approx\left[h_z v_c(r) / r\right]^2 + \sigma_0^2$. The value of $\sigma_r(1.8 R_e)$, from which $\sigma_0$ is derived, is assigned following the redshift evolution measured by \citet{Ubler_19}.
    
        For each realisation, we draw a random position and sky orientation within $1'$ of the brightest cluster galaxy in Abell 370, and construct a $25~\AA{} \times 6.4'' \times 6.4''$ ($20 \times 32 \times 32~\mathrm{px}$) synthetic MUSE datacube assuming the lensing model of \citet{Lagattuta_19}. The mock galaxy is centred in the cube and assigned a random inclination drawn from a uniform distribution in $\cos(i)$, restricted to $i > 30^\circ$, and oriented at a near-diagonal angle relative to the cube axes. Its SFR is drawn from the star-forming main sequence of \citet{Boogaard_18} and converted to a total \OII{} luminosity using the calibration of \citet{Gilbank_10}:
        \begin{equation}
        \begin{aligned}
            L_\mathrm{[OII]} = \mathrm{SFR} {} & \times \left( a \times \tanh\left[(x-b)/c \right] + d \right) \\
                                               & \times 1.06 \times \qty{2.53e40}{\erg.\s^{-1}},
        \end{aligned}
        \end{equation}
        where $L_\mathrm{[OII]}$ is the luminosity of the \OII{} doublet in \qty{}{\erg.\s^{-1}}, $x = \log \left( M_\star / M_\odot \right)$, $(a, b, c, d) = (-1.424, 9.827, 0.572, 1.700)$, and the factor $1.06$ is used to convert the \citet{Gilbank_10} calibration from a \citet{Kroupa_01} IMF to a \citet{Chabrier_03} IMF assumed in \citet{Boogaard_18}. The \OII{} emission is modelled as a double Gaussian with a fixed flux ratio $r^{\mathrm{[OII]}} = F^{\lambda3727} / F^{\lambda3729} = 0.7$.
        
        We generate a datacube free from instrumental resolution with an initial sampling of $0.04'' \times 0.04''$ (that is, oversampled by a factor of 5), and use it to compute the luminosity-weighted magnification. To reproduce the observed magnification distribution (see Fig.~10 of \citetalias{Richard_21}), we accept or reject mocks based on a probability density function proportional to $\mu^{-2}$ and imposing $1.5 < \mu < 6$. The lower limit accounts for the limited spatial coverage of the MUSE data, while the upper limit ensures that mock galaxies remain within a reasonable datacube size.
    
        For each accepted galaxy, the datacube is convolved with the PSF model from \citetalias{Richard_21}, the Gaussian LSF model from \citet{Bacon_17}, and resampled to the MUSE spatial sampling ($0.2'' \times 0.2''$). Gaussian white noise is then added to each pixel to match the surface-brightness limit corresponding to a 2-hour exposure time in \citetalias{Richard_21}. We generate 3000 mock galaxies following this procedure, and measure the maximum S/N ($\mathrm{S/N}_\mathrm{max}$) by fitting the \OII{} doublet in the central spaxel. We retain only galaxies with an effective S/N (introduced in Sect.~\ref{subsect:recovery_perf}) $\mathrm{S/N}_\mathrm{eff} = \mathrm{S/N}_\mathrm{max} \times \sqrt{\mu} R_e / R_\mathrm{PSF} > 5$, where $R_\mathrm{PSF}$ is the PSF half width at half maximum, and satisfying $v_{1.8}/\sigma_0 > 1$. This selection yields a final sample of 1923 mock galaxies.

    \subsection{Recovery performance}
    \label{subsect:recovery_perf}
        We extract the kinematics of our mock galaxies twice: first using a kinematic model similar to the one employed for its generation, but with an oversampling factor of 2 instead of 5 to reduce computation time; and second, treating the galaxy as if it were not lensed, with the resulting parameters simply rescaled by the magnification factor where relevant. In both cases, we assume that the true PSF and LSF are known without error\footnote{The impact of PSF FWHM errors is discussed in Sect.~4.4 of \citet{Bouche_15b}.}. The adopted priors are summarised in Tab.~\ref{tab:kinematic_priors}.

        \begin{table}[]
            \caption{Main parameters of our kinematic model, together with their adopted priors or fixed values.}
            \label{tab:kinematic_priors}
            \begin{tabularx}{\hsize}{@{\extracolsep{\fill}} lc}
                \hline
                \hline
                \noalign{\smallskip}
                Parameter & Prior / Value \\
                \noalign{\smallskip}
                \hline
                \noalign{\smallskip}
                \noalign{\smallskip}
                Sérsic index $n$ & $\mathrm{\mathcal{U}\left[0.5, 4.0\right]}$ \\
                \noalign{\smallskip}
                Effective radius $R_e \; [\mathrm{kpc}]$ & $\mathrm{\mathcal{U}\left[0.3, 30\right]}$ \\
                \noalign{\smallskip}
                Scale height $h_z / \sqrt{2\ln(2)}$ & $0.13 \times R_e$ \\
                \noalign{\smallskip}
                Inclination $i \; [\mathrm{deg}]$ & $\mathcal{N}_\mathrm{tr}\left(i^\mathrm{true}, 5, \pm 15\right)$ \\
                \noalign{\smallskip}
                Asymptotic velocity $v_\mathrm{max} \; [\mathrm{km/s}]$ & $\mathrm{\mathcal{U}\left[10, 500\right]}$ \\
                \noalign{\smallskip}
                Turnover radius $r_t \; [\mathrm{kpc}]$ & $\mathrm{\mathcal{U}\left[0.01, 3\right]}$ \\
                \noalign{\smallskip}
                Velocity dispersion floor $\sigma_0 \; [\mathrm{km/s}]$ & $\mathrm{\mathcal{U}\left[0, 100\right]}$ \\
                \noalign{\smallskip}
                Doublet ratio $r^\mathrm{[OII]}$ & $\mathrm{\mathcal{U}\left[0.35, 1.5\right]}$ \\
                \noalign{\smallskip}
                \hline
            \end{tabularx}
            \tablefoot{Other parameters (position, flux and PA) are assigned broad uniform priors, following the default settings in \galpak{}. A truncated Gaussian prior is denoted as $\mathcal{N}_\mathrm{tr}(x, y, z)$, with centre $x$, parent standard deviation $y$, and truncation bounds $z$. A uniform prior over the interval $\left[x, y\right]$ is written as $\mathcal{U}\left[x, y\right]$.}
        \end{table}

        \begin{figure*}[ht!]
            \includegraphics[width=\textwidth]{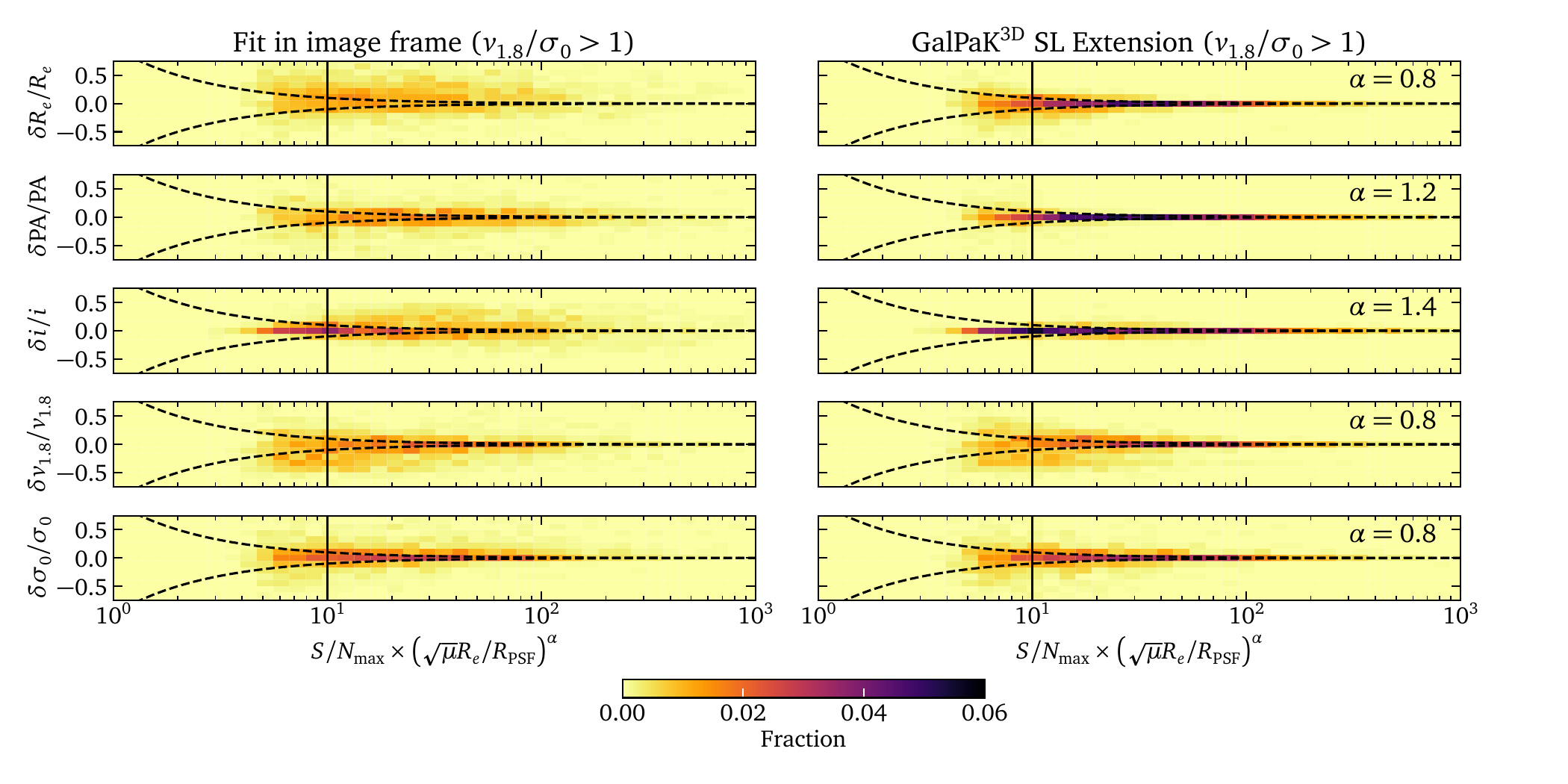}
            \caption{Comparison of the relative errors in the inferred properties, defined as $\delta p / p = (p^\mathrm{fit} -  p^\mathrm{true}) / p^\mathrm{true}$, for a kinematic model fitted either directly in the image frame (left column) or in the source frame using the \galpak{} Strong Lensing Extension (right column). Each row shows the error density distribution for a given property, restricted to the rotation-dominated subsample of mock galaxies ($v_{1.8}/\sigma_0 > 1$). The effective radius is corrected a posteriori by the magnification when fitted in the image plane. The PA error is relative to $\mathrm{PA}^\mathrm{true} = \qty{130}{\degree}$, as in \citet{Bouche_15b}. Dashed lines show the empirical relation $\delta p / p = \pm S/N_\mathrm{max} \times (\sqrt{\mu} R_e /R_\mathrm{PSF})^{\alpha}$ based on the coefficients $\alpha$ from \citet{Bouche_15b}, while solid vertical lines indicate $S/N_\mathrm{max} \times (\sqrt{\mu} R_e /R_\mathrm{PSF})^{\alpha} = 10$.}
            \label{fig:mock_relative_error}
        \end{figure*}

        \begin{figure*}[ht!]
            \includegraphics[width=\textwidth]{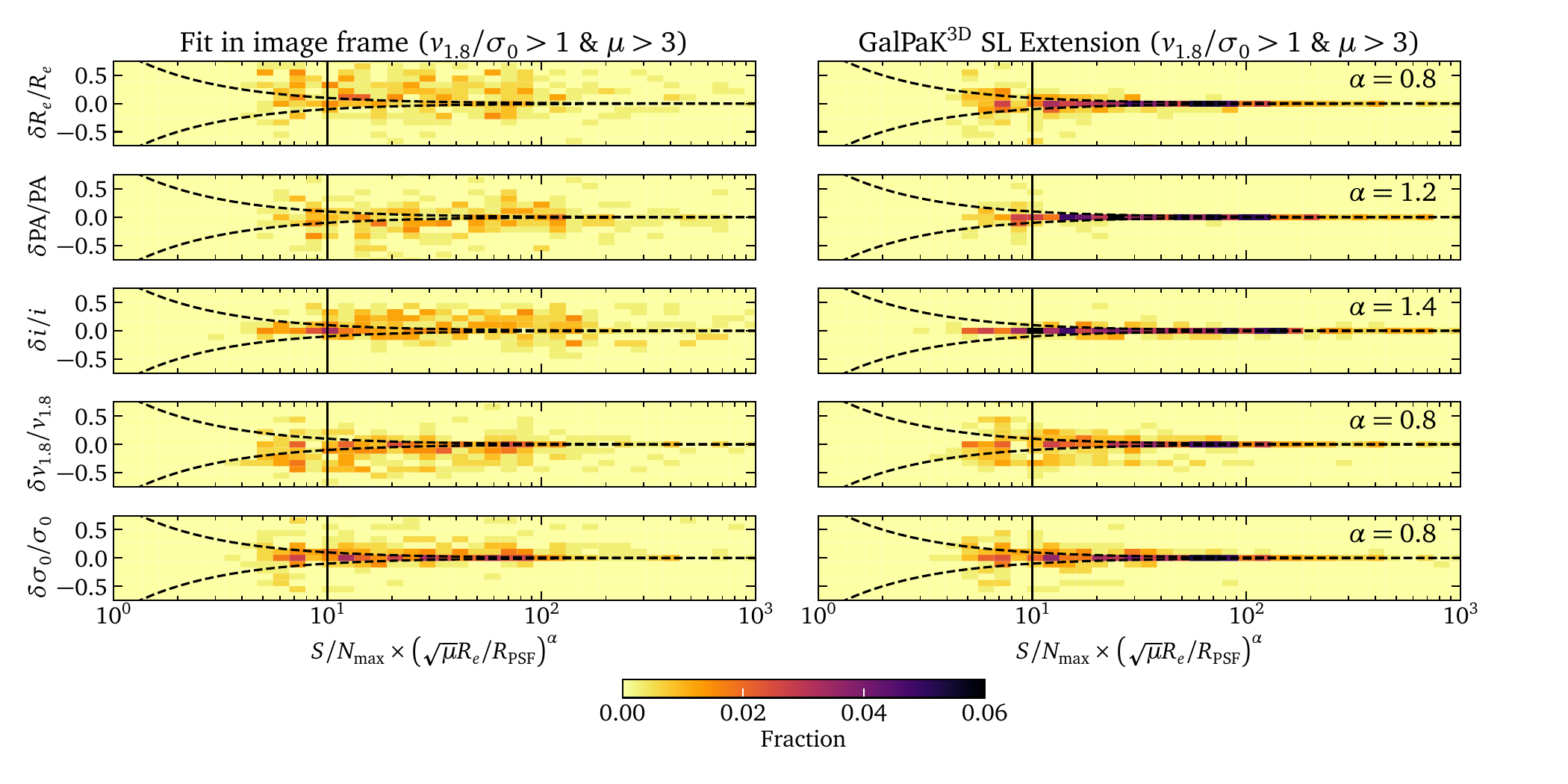}
            \caption{Same as Fig.~\ref{fig:mock_relative_error}, additionally restricted to the subsample of mock galaxies satisfying $\mu > 3$.}
            \label{fig:mock_relative_error_highmag}
        \end{figure*}

        Fig.~\ref{fig:mock_relative_error} shows the relative errors in the recovered parameters, defined as $\delta p / p = (p^\mathrm{fit} - p^\mathrm{true}) / p^\mathrm{true}$. Fig.~\ref{fig:mock_relative_error_highmag} presents the same quantities but is restricted to mock galaxies with magnifications $\mu > 3$. In both figures, the relative errors are plotted as a function of $\mathrm{S/N}_\mathrm{max}$ scaled by a power $\alpha$ of the apparent size-to-PSF ratio $\sqrt{\mu} R_e /R_\mathrm{PSF}$.

        In the marginally resolved regime ($R_e / R_\mathrm{PSF} \simeq 1$), the accuracy of recovered properties is known to depend not only on the S/N but also on the compactness of the source relative to the PSF. For blank-field galaxies, \citet{Bouche_15b} showed that the relative uncertainties in $R_e$, PA, $i$, and $v_\mathrm{max}$ scale inversely with both the central surface brightness (or S/N) and $(R_e / R_\mathrm{PSF})^{\alpha}$, with $\alpha \simeq 0.8$, $1.2$, $1.4$, and $0.8$, respectively. They therefore introduced an effective S/N, $\mathrm{S/N}_\mathrm{eff} = \mathrm{S/N}_\mathrm{max} \times (R_e / R_\mathrm{PSF})$, to capture this dependence.

        When modelling our mock galaxies with the \galpak{} Strong Lensing Extension, we recover similar trends\footnote{For $v_{1.8}$ and $\sigma_0$, we suppose $\alpha = 0.8$ based on the coefficient derived by \citet{Bouche_15b} for $v_\mathrm{max}$.} with $\mathrm{S/N}_\mathrm{max}$ and apparent size-to-PSF ratio (indicated by the dashed lines in the figures). This indicates that, once differential magnification is properly accounted for, the relative errors scale approximately as $\mu^{-\alpha/2}$ within the magnification range explored here. In contrast, this behaviour is much less apparent when fitting directly in the image plane, particularly for the higher-magnification subsample shown in Fig.~\ref{fig:mock_relative_error_highmag}. This suggests that neglecting differential magnification introduces a limiting source of error that cannot be mitigated by increased S/N alone.

        For galaxies satisfying $S/N_\mathrm{max} \times (\sqrt{\mu} R_e / R_\mathrm{PSF})^{\alpha} > 10$, both approaches yield negligible median deviations (within $5\%$) for most parameters. The main exception is the effective radius, which is biased by $+9\%$ when fitted directly in the image plane. In terms of relative error, ignoring (accounting for) differential magnification results in standard deviations of $31\%$ ($8\%$) for $R_e$, \qty{59}{\degree} (\qty{49}{\degree}) for the PA, $16\%$ ($4\%$) for the inclination, $18\%$ ($14\%$) for $v_{1.8}$, and $25\%$ ($15\%$) for $\sigma_0$.

        The benefits of including lensing deflections in the forward modelling become more pronounced at higher magnifications. Restricting the analysis to $3 < \mu < 6$, we find that direct image-plane fitting leads to significant median deviations even for objects with $S/N_\mathrm{max} \times (\sqrt{\mu} R_e / R_\mathrm{PSF})^{\alpha} > 10$, namely $+14\%$ for $R_e$, $+6\%$ for the inclination, and $-7\%$ for $v_{1.8}$. The corresponding standard deviations in the relative errors increase to $44\%$ for $R_e$, \qty{70}{\degree} for the PA, $18\%$ for the inclination, $20\%$ for $v_{1.8}$, and $38\%$ for $\sigma_0$ when differential magnification is ignored, compared to $10\%$, \qty{50}{\degree}, $5\%$, $14\%$, and $11\%$, respectively, when it is accounted for.

        Finally, we restrict the sample to mock galaxies with apparently well-constrained velocities for both fitting approaches, defined by $\Delta v_{1.8}^\mathrm{fit} / v_{1.8}^\mathrm{fit} < 30\%$, where $\Delta v_{1.8}$ is the posterior standard deviation. Even in this case, fitting directly in the image plane yields a significant median deviation of $+10\%$ in $R_e$. The corresponding standard deviations in the relative errors, ignoring (accounting for) differential magnification, are $28\%$ ($7\%$) for $R_e$, \qty{27}{\degree} (\qty{7}{\degree}) for the PA, $17\%$ ($4\%$) for the inclination, $14\%$ ($8\%$) for $v_{1.8}$, and $27\%$ ($16\%$) for $\sigma_0$.

        Overall, those validation tests demonstrate that the intrinsic morpho-kinematic properties of strongly lensed SFGs can be reliably recovered for rotation dominated ($v_{1.8}/\sigma_0 > 1$) galaxies which satisfy $S/N_\mathrm{max} \times (\sqrt{\mu} R_e / R_\mathrm{PSF})^{\alpha} \gtrsim 10$, with $\alpha \simeq 1$. We note that the recovered parameters exhibit relative uncertainties comparable to those obtained in the unlensed case at similar $\mathrm{S/N}_\mathrm{eff}$, when differential magnification is accounted for. We find that this approach is significantly more robust, particularly for morphological parameters and velocity dispersion, with improvements of factors $\simeq 2-4$ in relative errors, than methods that neglect differential magnification, even for moderate magnifications ($\mu < 6$). Circular velocity also benefits from this approach, albeit to a lesser extent, as the improved inclination recovery entails a more accurate deprojection of the observed velocity field.

\section{A robust sample of strongly-lensed SFGs observed with MUSE}
\label{sect:sample}

    \subsection{Sample definition}
        Our sample selection is guided by data quality and by the ability of our morpho-kinematic modelling approach to robustly recover source properties. As shown in Sect.~\ref{sect:recovery_perf}, the relative error of recovered parameters scales approximately with $\mathrm{S/N}_\mathrm{eff}^{-1}$. In the following paragraphs, we detail how it is measured and present our sample selection, which consists of a series of cuts based on $\mathrm{S/N}_\mathrm{eff}$ and other observational properties. We also present complementary measurements (e.g. single and double-Sérsic modelling) to better characterise the sample and to assess the consistency of our morpho-kinematic results relative to parameters inferred from broadband imaging.
        
        We start by extracting a base sample of \OII{}$\lambda\lambda3727, 3729$ emitters from the MUSE Lensing Cluster catalogues of Abell 370, Abell 2744, MACS0416 and Abell S1063, retaining only sources with a secure redshift estimate (\texttt{ZCONF = 3}). The spectral coverage of MUSE limits the redshift range, which we further restrict to exclude cluster members ($z = 0.3 - 0.5$). The resulting sample comprises galaxies with $0.5 < z < 1.5$. We then match this sample with the photometric catalogue of \citet{Shipley_18} and visually inspect the detections to remove duplicates from segmentation errors, galaxies located at the edges of the MUSE mosaic, galaxy–galaxy lensed systems, and perturbed galaxies showing tidal features.
        
        \subsubsection{Lensing magnification and morphological properties}
        \label{subsubsect:magnification_and_morphology}
            For each galaxy, we measure the gravitational lensing magnification and intrinsic structural parameters within a manually defined region chosen to prevent contamination from neighbouring galaxies. Deflection and magnification maps are generated from the cluster mass models referenced in Tab.~\ref{tab:clusters}, and the galaxy magnification is computed as the harmonic mean within the useful region.
    
            We then forward-model the H-band (HST/WFC3 F160W, rest-frame I band) flux distribution using the \texttt{cleanlens} mode of the \texttt{Lenstool} software. The intrinsic surface brightness is modelled as a single \citet{Sersic_63} profile with seven free parameters: centre position, magnitude, scale radius, Sérsic index $n$, PA, and ellipticity $\epsilon$. The model flux distribution is lensed with the corresponding lensing mass model, convolved with the empirical F160W PSF from \citet{Shipley_18}, and combined with a sky background estimated via $3\sigma$-clipping to produce a model image-frame flux distribution. MCMC sampling is used to compare the modelled and observed distributions, and to obtain the posterior distributions of the model parameters assuming broad uniform priors. From these, we derive the intrinsic effective radius and the source-frame axis ratio $q^\mathrm{src} = (1 - \epsilon) / (1 + \epsilon)$. We compute the inclination $i$, assuming $\cos^2(i) = (q^2 - q_0^2) / (1 - q_0^2)$ with an intrinsic stellar disc axis ratio $q_0 = 0.25$ \citep{Lian_24, Tsukui_25}. For visualisation, we also perform a direct source reconstruction by mapping image pixels onto a regular source pixel grid.

        \subsubsection{S/N and kinematics of the \OII{} line}
        \label{subsubsect:kinematic_maps}
            To prepare the MUSE data for kinematic analysis, we extract $100$~\AA{} sub-datacubes centred on the \OII{} doublet, using the positions and spectroscopic redshifts from \citetalias{Richard_21}. We mask the central $25$~\AA{} and fit the continuum in each spaxel as a linear ramp between the mean flux blueward and redward of the doublet. After subtracting the continuum, we crop the sub-datacubes to the central $25$~\AA{}.
    
           We then derive kinematic maps to visually inspect the velocity fields. We adaptively bin spaxels to a target S/N of $\simeq 5$ using the \texttt{PowerBin} algorithm \citep{Cappellari_25}. To ensure that the binning maximises the S/N in the outer regions, we first retain only spaxels with S/N $> 2$, where the S/N is estimated from a \OII{} pseudo–narrowband image smoothed with a Gaussian kernel ($\sigma = 1$ spaxel). In each resulting bin, the \OII{} doublet is modelled with two Gaussian components sharing a common velocity and velocity dispersion. The flux, velocity, and velocity dispersion are inferred using Bayesian sampling with the affine-invariant MCMC ensemble sampler \texttt{emcee} \citep{Foreman-Mackey_13}, using 10 walkers run for 10000 steps, including a burn-in of 2000 steps. For reference, we also fit the \OII{} doublet in the brightest individual spaxel to estimate $\mathrm{S/N}_\mathrm{max}$.

        \subsubsection{Selection criteria}
            Our final sample is defined by the following criteria:
            \begin{enumerate}
                \item Secure spectroscopic redshift (\texttt{ZCONF = 3}) with $0.5 < z < 1.5$;
                \item No evidence of gravitational perturbations (e.g. tidal arms), edge effects, or contamination, based on visual inspection.
            \end{enumerate}
            For robust kinematic recovery, we additionally require:
            \begin{enumerate}
                \setcounter{enumi}{2}
                \item $\mathrm{S/N}_\mathrm{eff} = \mathrm{S/N}_\mathrm{max} \times \sqrt{\mu} R_e /R_\mathrm{PSF} > 10$
                \item $\sqrt{\mu} R_e /R_\mathrm{PSF} > 1/2$, that is, the stellar disc must span at least one resolution element;
                \item $i > 30~\mathrm{deg}$, excluding face-on galaxies.
            \end{enumerate}

            \begin{figure}[]
                \includegraphics[width=\hsize]{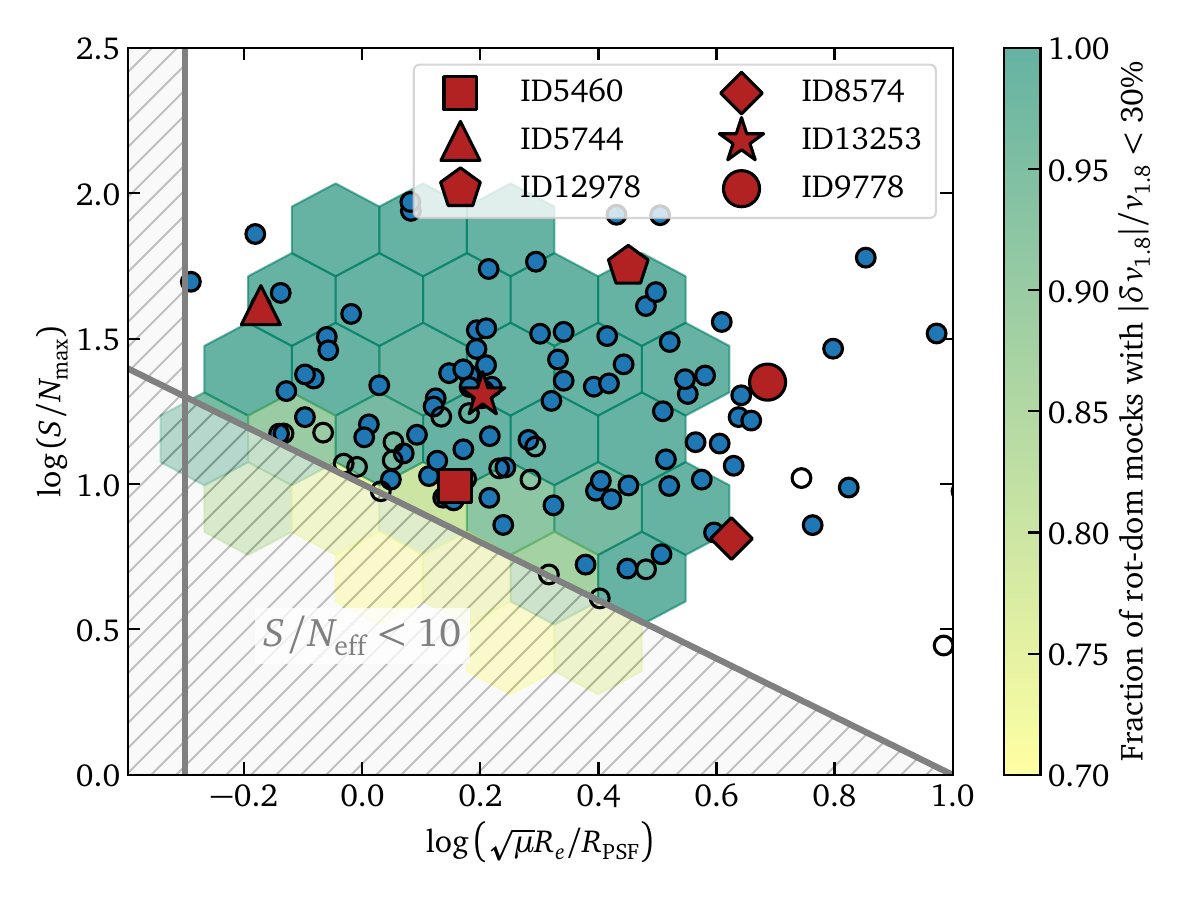}
                \caption{S/N of the brightest \OII{} spaxel against the apparent size-to-PSF ratio. The hatched grey region marks the parameter space excluded by our selection. Red symbols mark 6 galaxies chosen to showcase the range of apparent sizes and S/N in Fig.~\ref{fig:highlights}. Open symbols mark galaxies that are either dispersion-dominated ($v_{1.8}/\sigma_0 \leq 1$) or have poorly constrained velocities ($\Delta v_{1.8}/v_{1.8} \geq 30\%$), where $\Delta v_{1.8}$ is the posterior standard deviation. The background hexagonal bins illustrate the fraction of rotation-dominated ($v_{1.8}/\sigma_{0} > 1$) mock galaxies with a robust velocity recovery ($|v^\mathrm{fit}_{1.8} - v^\mathrm{true}_{1.8}| / v^\mathrm{true}_{1.8} < 30\%$). Bins with less than 20 mocks are not shown.}
                \label{fig:selection}
            \end{figure}

            Our selection (Fig.~\ref{fig:selection}) comprises \ngal{} galaxies excluding 8 counter-images of galaxies already in the sample. For these cases where two counter-images of the same galaxy pass our selection criteria, we exclude the counter-image with the lower $\mathrm{S/N}_\mathrm{eff}$ from the final sample.

    \subsection{Additional measurements}
    
        \subsubsection{Bulge-to-total flux ratio}
        \label{subsubsect:BT}
            We estimate the bulge-to-total flux ratio ($\mathrm{B/T}_\mathrm{flux}$) by re-modelling the H-band flux distribution with a two-component bulge–disc model, following the general steps of Sect.~\ref{subsubsect:magnification_and_morphology}. We adopt an exponential disc ($n_d = 1$) plus a de Vaucouleurs bulge ($n_b = 4$), with shared centre, PA, and ellipticity. Initial guesses and constraints are drawn from the single-Sérsic fit. Notably, the disc (bulge) effective radius is forced to be larger (smaller) than the single-component effective radius. This model has eight free parameters: common centre, PA, and ellipticity, plus distinct magnitudes and scale radii for the bulge and disc. We infer $\mathrm{B/T}_\mathrm{flux}$ from the posterior probability distribution samples.

        \subsubsection{Stellar population properties}   
            Stellar population properties are derived from spectral energy distribution (SED) modelling using the photometry of \citet{Shipley_18}. Their catalogues provide fluxes corrected to total fluxes based on the ratio of \texttt{SExtractor}’s \texttt{AUTO} flux to an aperture flux in the F160W filter. To ensure stellar masses and SFHs consistent with our morphology and kinematics measurements, we apply an additional aperture correction: for each band, the \citet{Shipley_18} flux is scaled by the ratio of the total F160W flux derived from the best-fit \texttt{Lenstool} bulge–disc model (Sect.~\ref{subsubsect:BT}) to the reported F160W flux, both similarly corrected for Galactic extinction. For galaxies segmented across multiple catalogue entries, we combine the corresponding fluxes and uncertainties before applying this correction.
            
            We then remove invalid flux values, correct the fluxes and uncertainties for lensing magnification, and impose a minimum relative uncertainty of $5\%$ except for Spitzer/IRAC channels where we impose $10\%$. This error floor, following \citet{Carnall_18, Carnall_19}, is intended to prevent underestimating uncertainties in the high-S/N regime, where calibration errors dominate the photometric error. We do not include magnification uncertainties at this stage, as doing so would unphysically allow the model spectral slope to vary within the magnification error range.
            
            For each galaxy, we fit the SED at the spectroscopic redshift estimate of \citetalias{Richard_21}, using the Bayesian spectral fitting code \texttt{BAGPIPES} \citep{Carnall_18}. Our model relies on the 2016 version of the \citet{BruzualCharlot_03} stellar population synthesis models with a \citet{Kroupa_01} IMF, a free metallicity and the \citet{Calzetti_00} reddening law. Following \citet{Leja_19a, Leja_19b}, we adopt a nonparametric SFH divided into seven bins of lookback time, with the SFR constant within each bin and linked by a continuity prior to disfavour sharp transitions between bins. The bins are defined at 0–30, 30–100, 100–330, 330–1100 Myr, and $0.85t_\mathrm{univ}$–$t_\mathrm{univ}$, where $t_\mathrm{univ}$ is the age of the Universe at the observed redshift. The remaining two bins are equally spaced in logarithmic time between 1100 Myr and $0.85t_\mathrm{univ}$. The model parameters and their priors are summarised in Tab.~\ref{tab:SED_priors}.
    
            \begin{table}[!h]
                \caption{Parameters of our SED model, together with their adopted priors or fixed values.}
                \label{tab:SED_priors}
                \begin{tabularx}{\hsize}{@{\extracolsep{\fill}} lc}
                    \hline
                    \hline
                    \noalign{\smallskip}
                    Parameter & Prior / Value \\
                    \noalign{\smallskip}
                    \hline
                    \noalign{\smallskip}
                    \noalign{\smallskip}
                    Stellar mass formed $\log \left( M_\mathrm{formed} / M_\odot \right) \; [\mathrm{dex}]$ & $\mathrm{\mathcal{U}\left[0, 13\right]}$ \\
                    \noalign{\smallskip}
                    Stellar metallicity $Z / Z_\odot$ & $\mathrm{\log\mathcal{U}\left[0.01, 5\right]}$ \\
                    \noalign{\smallskip}
                    $\log \left( \mathrm{SFR}_n / \mathrm{SFR}_{n+1}\right)$ between bin $n$ and $n+1$ & $t_\nu(0.3, 2)$ \\
                    \noalign{\smallskip}
                    Dust attenuation in the V band $A_V\; [\mathrm{mag}]$ & $\mathrm{\mathcal{U}\left[0, 4\right]}$ \\
                    \noalign{\smallskip}
                    Multiplicative factor on $A_V$ & \multirow{2}{*}{$3$} \\
                    for stars in birth clouds $\epsilon$ & \\
                    \noalign{\smallskip}
                    Maximum age of birth clouds $a_\mathrm{BC} \; [\mathrm{Myr}]$ & $10$ \\
                    \noalign{\smallskip}
                    Ionization parameter $\log U$ & $-3$ \\
                    \noalign{\smallskip}
                    Redshift $z$ & $z_\mathrm{spec}^\mathrm{R21}$ \\
                    \noalign{\smallskip}
                    \hline
                \end{tabularx}
                \tablefoot{A uniform prior over the interval $\left[x, y\right]$ is denoted $\mathcal{U}\left[x, y\right]$, or $\log \mathcal{U}\left[x, y\right]$ if uniform in log-space, and $t_\nu(\sigma, \nu)$ indicates a Student-t distribution with scale parameter $\sigma$ and $\nu$ degrees of freedom.}
            \end{table}

            \begin{figure*}[]
                \includegraphics[width=\textwidth]{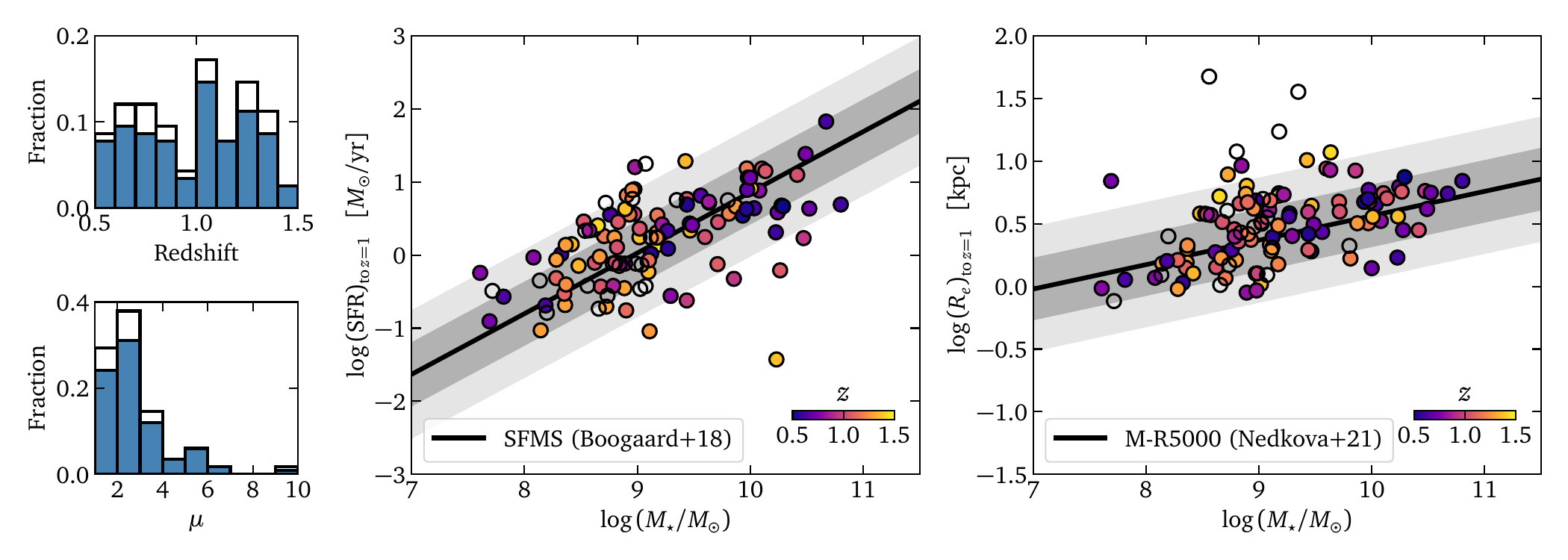}
                \caption{General sample properties. The panels show the distributions of redshift (top left) and magnification (bottom left), together with the locations of the galaxies in the $M_\star$–SFR (middle) and $M_\star$–$R_e$ (right) planes. Effective radii are corrected to rest-frame 5000~\AA{} following Eqs. 1 and 2 of \citet{Wel_14}, expressed relative to the size–mass relation of \citet{Nedkova_21} (interpolated with redshift), and scaled to $z = 1$. Similarly, SFRs are expressed relative to the star-forming main sequence of \citet{Boogaard_18}, adjusted to the redshift and stellar mass of each galaxy, and normalized to $z = 1$. The shaded regions indicate the $\pm1\sigma$ (dark grey) and $\pm2\sigma$ (light grey) scatter around the corresponding relations. Open symbols and histogram segments mark galaxies that are either dispersion-dominated ($v_{1.8}/\sigma_0 \leq 1$) or have poorly constrained velocities ($\Delta v_{1.8}/v_{1.8} \geq 30\%$), where $\Delta v_{1.8}$ is the posterior standard deviation.}
                \label{fig:properties}
            \end{figure*}

            We apply corrections of $-0.034~\mathrm{dex}$ and $-0.027~\mathrm{dex}$ to convert our stellar masses and SFRs, respectively, to a \citet{Chabrier_03} IMF \citep{Madau_14}.

    \subsection{General sample properties}
        
        \begin{figure*}[]
            \includegraphics[width=\textwidth]{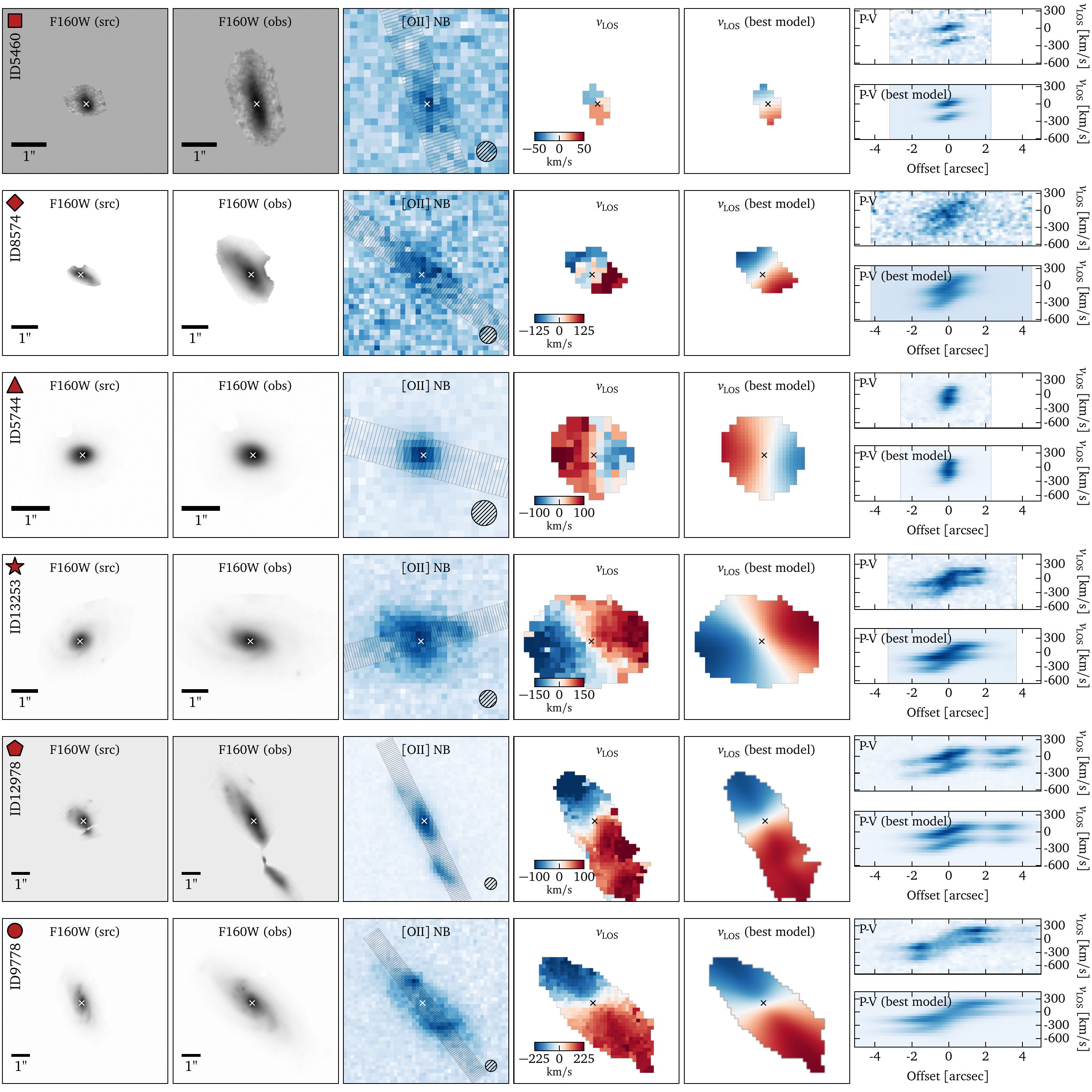}
            \caption{For each galaxy highlighted in Fig.~\ref{fig:selection}, we present the reconstructed HST/F160W source-plane image, the corresponding observed HST/F160W image, the MUSE \OII{} pseudo-narrowband, the LOS velocity map and best-fitting \galpak{} model, as well as P-V diagrams extracted from both the observed and model datacubes. The MUSE PSF FWHM is indicated by the hatched circle. The sequence of grey rectangles marks the 1 arcsec-wide slit used to extract the P-V diagrams.}
            \label{fig:highlights}
        \end{figure*}
        
        The general properties of our sample (Fig.~\ref{fig:properties}) are broadly representative of main-sequence SFGs over $0.56 \le z \le 1.37$ and $8.1 \le \log \left( M_\star / M_\odot \right) \le 10.3$, observed under moderate magnification in the range $1.4 \le \mu \le 12.4$ (5\textsuperscript{th}–95\textsuperscript{th} percentiles). The lower tail of the effective radius distribution is under-represented in our selection, leading to an overall $+0.08~\mathrm{dex}$ offset relative to the size-mass relation of \citet{Nedkova_21}. In terms of structural parameters, we note that the majority of our sample is disc-dominated with 70\% of galaxies having $\mathrm{B/T}_\mathrm{flux} < 0.3$. This disc-dominated subset has nearly exponential light profiles as observed in the H band (rest-frame I band), with a median Sérsic index of 1.0. We compare the properties of the mock sample and the kinematic sample in Fig.~\ref{fig:mock_properties}.

\section{Derivation of the TFR at $z \sim 1$}
\label{sect:results}

    \subsection{Morphology and kinematics of the sample}
    
        We extract the kinematics of our sample using the model described in Sect.~\ref{sect:recovery_perf} and the main priors listed in Tab.~\ref{tab:kinematic_priors}. The model comprises 12 free parameters, $\bm{\theta} = \{\mathbf{r}_0^\mathrm{img}, z_0^\lambda, F^\mathrm{[OII]}, R_e, n, i, \mathrm{PA}, \sigma_0, r_t, v_\mathrm{max}, r^\mathrm{[OII]}\}$, which jointly describe the apparent position and flux, together with the intrinsic structural properties and kinematics (corrected for pressure support). For the inclination, we adopt a Gaussian prior centred on the single-Sérsic H-band estimate ($i^\mathrm{F160W}$, see Sect.~\ref{subsubsect:magnification_and_morphology}), with a scatter of $5~\mathrm{deg}$ and truncated at $\pm 15~\mathrm{deg}$. We additionally restrict the centre position to $\pm 1~\mathrm{spaxel}$ around the H-band two-component bulge-disc model. We use the deflection maps and Moffat PSF from \citetalias{Richard_21}, and model the LSF as a wavelength-dependent Gaussian following \citet{Bacon_17}. While our extension does not account for uncertainties in the cluster mass model and the resulting deflection maps, Appendix~\ref{appendix:mass_modelling} shows that using an average mass model built from five publicly available models does not significantly affect the recovered velocities. In the following, we restrict our analysis to 95 galaxies that are rotation-dominated ($v_{1.8}/\sigma_0 > 1$) with well-constrained velocities ($\Delta v_{1.8} / v_{1.8} < 30\%$, where $\Delta v_{1.8}$ is the posterior standard deviation).
        
        In Fig.~\ref{fig:highlights}, we present the morphology and kinematics of galaxies chosen to span the range of size-to-PSF ratios and S/N highlighted in Fig.~\ref{fig:selection}. The first two columns display the HST/F160W source-plane reconstruction and the corresponding HST/F160W cutout. The third column shows the MUSE \OII{} pseudo-narrowband image. The fourth and fifth columns show, respectively, the observed (Sect.~\ref{subsubsect:kinematic_maps}) and the best-model velocity maps. The final column shows position-velocity (P-V) diagrams extracted with \texttt{pvextractor} \citep{Ginsburg_16}, using a 1 arcsec-wide slit oriented to maximise the velocity gradient in the best-fit model map. The apparent double structure in the P-V diagrams reflects the small wavelength separation of the two lines forming the \OII{} doublet, which manifests as two nearby components in velocity space. Those examples illustrate that the models are able to reproduce the lensed data, including cases such as ID13253 and ID9778 where differential magnification causes the apparent major and minor velocity axes to deviate from orthogonality, and ID12978 where the receding side is strongly distorted. They also highlight that, thanks to lensing magnification, part of our sample has sufficiently extended coverage to probe RCs beyond their turnover.

        Before modelling the TFR, we assess the consistency between structural parameters derived from the \OII{} datacubes and those obtained from broadband continuum imaging, finding good overall agreement. The median misalignment between the photometric and kinematic PA is \qty{6}{\degree}, and the median absolute deviation in $\log(R_e)$ is $0.14~\mathrm{dex}$, with a modest systematic offset of $-0.09~\mathrm{dex}$ in \OII{} relative to the single-Sérsic broadband estimate. Finally, we find that a substantial fraction of galaxies have \OII{} Sérsic indices close to the imposed lower bound of 0.5, yielding both a median value and a median absolute deviation of 0.6 for this parameter.

    \subsection{Velocity and mass definitions}
    
        In the following, we model the TFR as:
        \begin{equation}
        \label{eq:TFR}
            \log\left( M \right) = a \log\left( v \right) + b,
        \end{equation}
        where $M$ represents either the stellar or the baryonic mass (in $M_\odot$) and $v$ is the circular velocity at $1.8$ or $2R_e$ (in \qty{}{\km.\s^{-1}}). These radii are selected to match the velocity definitions adopted in local reference TFRs, and probe the outer, approximately flat region of the RCs.
        
        Baryonic masses are obtained by adding a cold gas component, estimated from scaling relations, to the stellar masses. Molecular gas masses ($M_\mathrm{gas,\,mol}$) follow the scaling relations of \citet[][their Tab.~2b]{Tacconi_20}. For the atomic gas, we use predictions from the \textsc{NeutralUniverseMachine} empirical model \citep{Guo_23}, which predicts the evolution of atomic and molecular gas based on observational constraints (including the HI-stellar mass relation at $z \sim 1$ from \citealt{Chowdhury_22}) and the \textsc{UniverseMachine} \citep{Behroozi_19} model for galaxy formation. Specifically, we adopt their best-fit relation between HI depletion timescale $\tau_\mathrm{HI}$ and stellar mass at $z \sim 1$\footnote{Using the \texttt{WebPlotDigitizer} \citep{WebPlotDigitizer}, we digitized the best-fit average relation between $\log \tau_\mathrm{HI} \equiv \log \langle M_\mathrm{HI} \rangle - \langle \log \mathrm{SFR} \rangle$ and $\langle \log M_\star \rangle$ from their Fig.~16, and converted it to a median relation assuming lognormal HI masses and the reported $0.8~\mathrm{dex}$ scatter in $\log \tau_\mathrm{HI}$.}, and use it to infer the median HI mass $\log M_\mathrm{HI, \, MS}(M_\star)$. We further account for the dependence of HI mass on offset from the SFMS using $\log(M_\mathrm{HI} / M_\mathrm{HI, \, MS}) = \lambda \log(\mathrm{SFR} / \mathrm{SFR}_\mathrm{MS})$ with $\lambda \approx 0.4$. The inferred HI mass is then multiplied by 1.33 to include helium \citep[e.g.][]{McGaugh_12}.

        The inferred atomic gas fraction, defined as $f_\mathrm{gas,\,at} = 1.33 M_\mathrm{HI} / M_\mathrm{bar}$, ranges between $<1\%$ and $77\%$ (5\textsuperscript{th}–95\textsuperscript{th} percentiles), with a median of $22\%$. For the molecular gas fraction, defined as $f_\mathrm{gas,\,mol} = M_\mathrm{gas,\,mol} / M_\mathrm{bar}$, the corresponding range is $13 - 53\%$, with a median of $34\%$.

    \subsection{Fitting procedure and robustness tests}
    
        We adopt a Bayesian approach and use the MCMC sampler \texttt{emcee} \citep{Foreman-Mackey_13} to obtain the posterior distributions of the model parameters. The fit follows an orthogonal maximum-likelihood approach \citep[e.g.][Appendix A]{Lelli_19}, assuming a constant Gaussian intrinsic scatter orthogonal to the best-fit relation ($\sigma_{\perp,\, \mathrm{int}}$) and accounting for independent uncertainties in velocity and mass. Velocity errors are taken as the posterior standard deviation from our kinematic fits, while stellar masses are assigned a uniform uncertainty of $\pm 0.15~\mathrm{dex}$ to capture modelling uncertainties, for example those related to the magnification, SFH, stellar evolution, nebular emission and dust attenuation assumptions \citep[$\sim 0.1~\mathrm{dex}$,][]{Pacifici_23}, which is larger than the median observational error reported by \texttt{BAGPIPES} ($0.06~\mathrm{dex}$). Similarly, baryonic masses are assigned a uniform uncertainty of $\pm 0.2~\mathrm{dex}$. Given the limited mass range of our sample, we fix the slope to local reference values in our fiducial analysis, and fit only for the zero-point and intrinsic scatter orthogonal to the relation assuming broad uniform priors. MCMC sampling is run with 20 walkers over 10000 steps, including a burn-in phase of 2000 steps.
        
        Investigations of TFR evolution depend on the adopted velocity definition, the choice of local reference and sample selection. In our fiducial analysis, we use $v_{1.8}$ for the sTFR and compare our results to \citet{Reyes_11}\footnote{To allow a meaningful comparison with the literature samples compiled by \citet{Turner_17}, we adopt their local reference relation derived from the \citet{Reyes_11} dataset.}. For the bTFR, we use $v_{2.0} = v_c(2R_e)$ and compare to \citet{Lelli_19}. In both cases, we retain only rotationally supported galaxies ($v / \sigma_0 > 1$) with well-constrained velocities ($\Delta v / v < 30\%$). We also perform fits with a free slope to verify that the local slopes remain compatible with our data. Finally, to evaluate the robustness of our results, we repeat the analysis, varying the velocity definition (at $1.8$ or $2R_e$), the rotational-support criterion ($v/\sigma_0 > 1$ or $v/\sigma_0 > 2$), and the adopted local reference (\citealt{Reyes_11} or \citealt{Ristea_24}).

    \begin{figure*}[!ht]
        \hspace*{\fill}
        \begin{minipage}[b]{0.33\textwidth}
            \centering
            \vspace{0pt}
            \includegraphics[width=\textwidth]{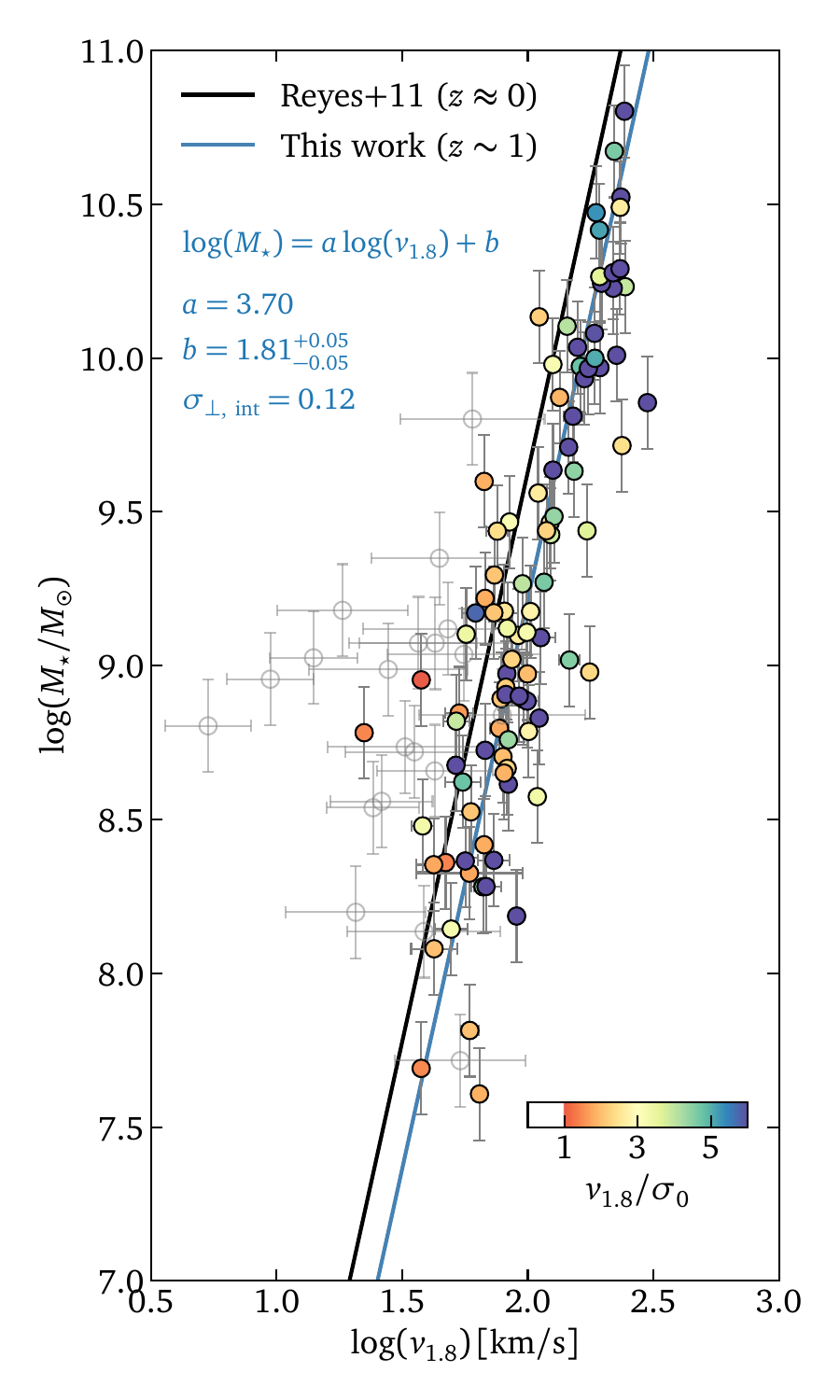}
        \end{minipage}
        \hfill
        \begin{minipage}[b]{0.33\textwidth}
            \centering
            \vspace{0pt}
            \includegraphics[width=\textwidth]{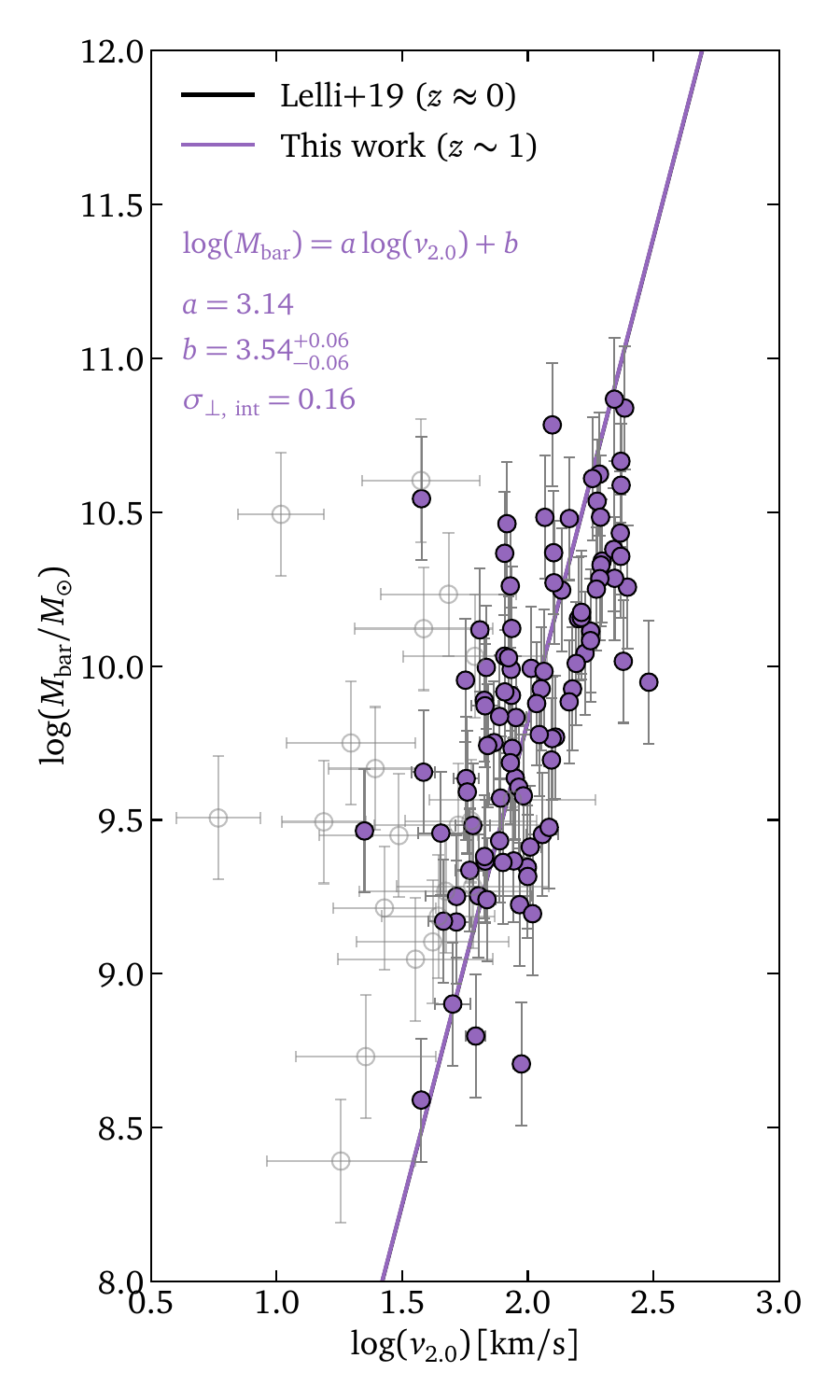}
        \end{minipage}
        \begin{minipage}[b]{0.32\textwidth}
            \centering
            \vspace{0pt}
            \includegraphics[width=\textwidth]{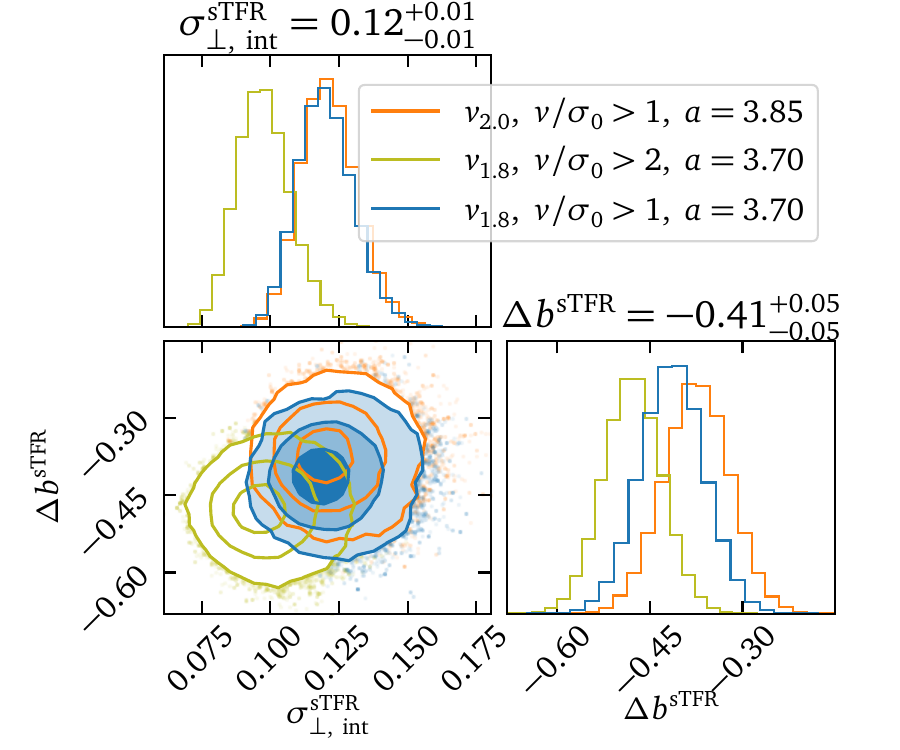}
            \includegraphics[width=\textwidth]{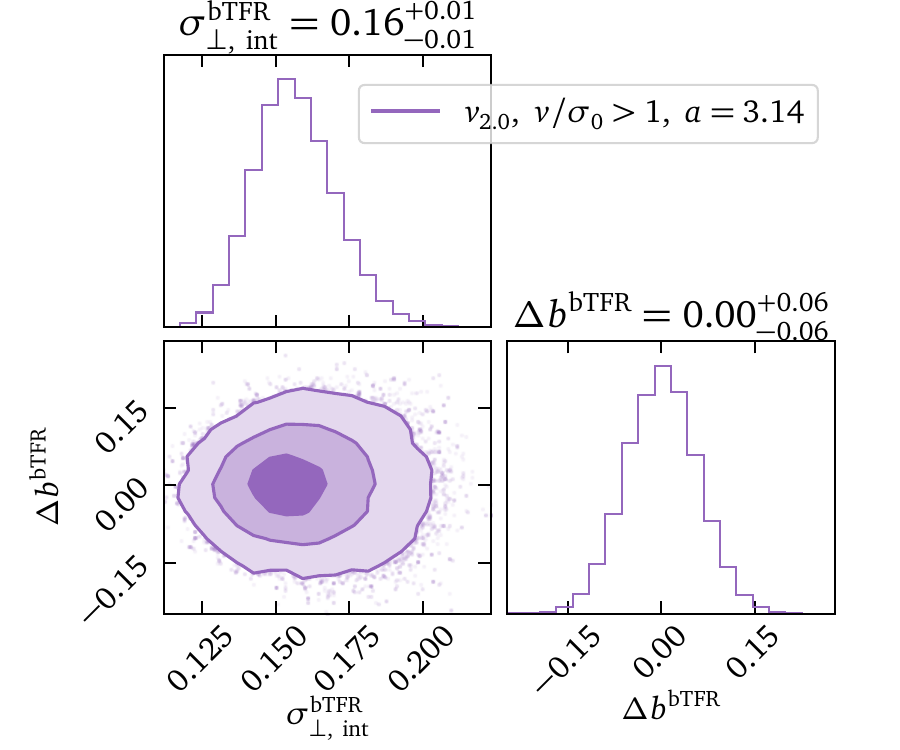}
        \end{minipage}
        \hspace*{\fill}

        \caption{sTFR (left) and bTFR (middle) for our sample. Solid blue and purple lines show the best-fit relations from our fiducial analyses, which adopt fixed slopes from the local references of \citet{Reyes_11} (sTFR) and \citet{Lelli_19} (bTFR), shown in black. Galaxies excluded from the fit (see text) are shown as open grey symbols. The top-right and bottom-right panels display the posterior distributions of these parameters for the sTFR and bTFR, respectively, with contours enclosing 39\%, 86\%, and 99\% of the samples. For the sTFR, filled contours correspond to the fiducial analysis, while open contours show results obtained when varying the sample selection (in green), or the velocity definition and adopted reference slope (in orange).}
        \label{fig:TFR}
    \end{figure*}

    \begin{table*}[!ht]
        \caption{TFR best-fit parameters.}
        \label{tab:TFR_fits}
        \begin{tabularx}{\hsize}{@{\extracolsep{\fill}} lcccccccc}
            \hline
            \hline
            \noalign{\smallskip}
             $M$ & $v$ & $v / \sigma_0$ & Local Reference & $a$ & $b_\mathrm{ref}$ & $b$ & $\sigma_{\perp\mathrm{,\ int}}$ & $N$ \\
            \noalign{\smallskip}
            \hline
            \noalign{\smallskip}
            \noalign{\smallskip}
            $M_\star$ & $v_{1.8}$ & $>1$ & free slope & $3.73^{+0.29}_{-0.26}$ & - & $1.75^{+0.51}_{-0.59}$ & $0.12$ & $95$\\
            \noalign{\smallskip}
            $M_\star$ & $v_{1.8}$ & $>1$ & \citet{Reyes_11} & $3.70$ & $2.22$ & $1.81^{+0.05}_{-0.05}$ & $0.12$ & $95$\\
            \noalign{\smallskip}
            $M_\star$ & $v_{1.8}$ & $>2$ & \citet{Reyes_11} & $3.70$ & $2.22$ & $1.74^{+0.05}_{-0.05}$ & $0.10$ & $79$ \\
            \noalign{\smallskip}
            $M_\star$ & $v_{2.0}$ & $>1$ & \citet{Ristea_24} & $3.85$ & $1.84$ & $1.47^{+0.05}_{-0.05}$ & $0.12$ & $95$ \\
            \noalign{\smallskip}
            \hline
            \noalign{\smallskip}
            $M_\mathrm{bar}$ & $v_{2.0}$ & $>1$ & free slope & $3.11^{+0.45}_{-0.36}$ & - & $3.61^{+0.72}_{-0.91}$ & $0.16$ & $95$ \\
            \noalign{\smallskip}
            $M_\mathrm{bar}$ & $v_{2.0}$ & $>1$ & \citet{Lelli_19} & $3.14$ & $3.54$ & $3.54^{+0.06}_{-0.06}$ & $0.16$ & $95$ \\
            \noalign{\smallskip}
            \hline
            \end{tabularx}
            \tablefoot{We test the robustness of the fits by varying the velocity definition (at $1.8$ or $2R_e$), the criterion to define rotationally supported galaxies ($v/\sigma_0 > 1$ or $v/\sigma_0 > 2$), and the adopted local reference slopes ($a$). For each configuration, we report the local reference zero-point ($b_\mathrm{ref}$), the best-fit zero-point ($b$), the orthogonal intrinsic scatter ($\sigma_{\perp,\ \mathrm{int}}$), and the number of galaxies entering the fit ($N$).}
    \end{table*}

\section{Evolution of the TFR since $z \sim 1$}
\label{sect:discussion}

    \begin{figure}[]
        \includegraphics[width=\hsize]{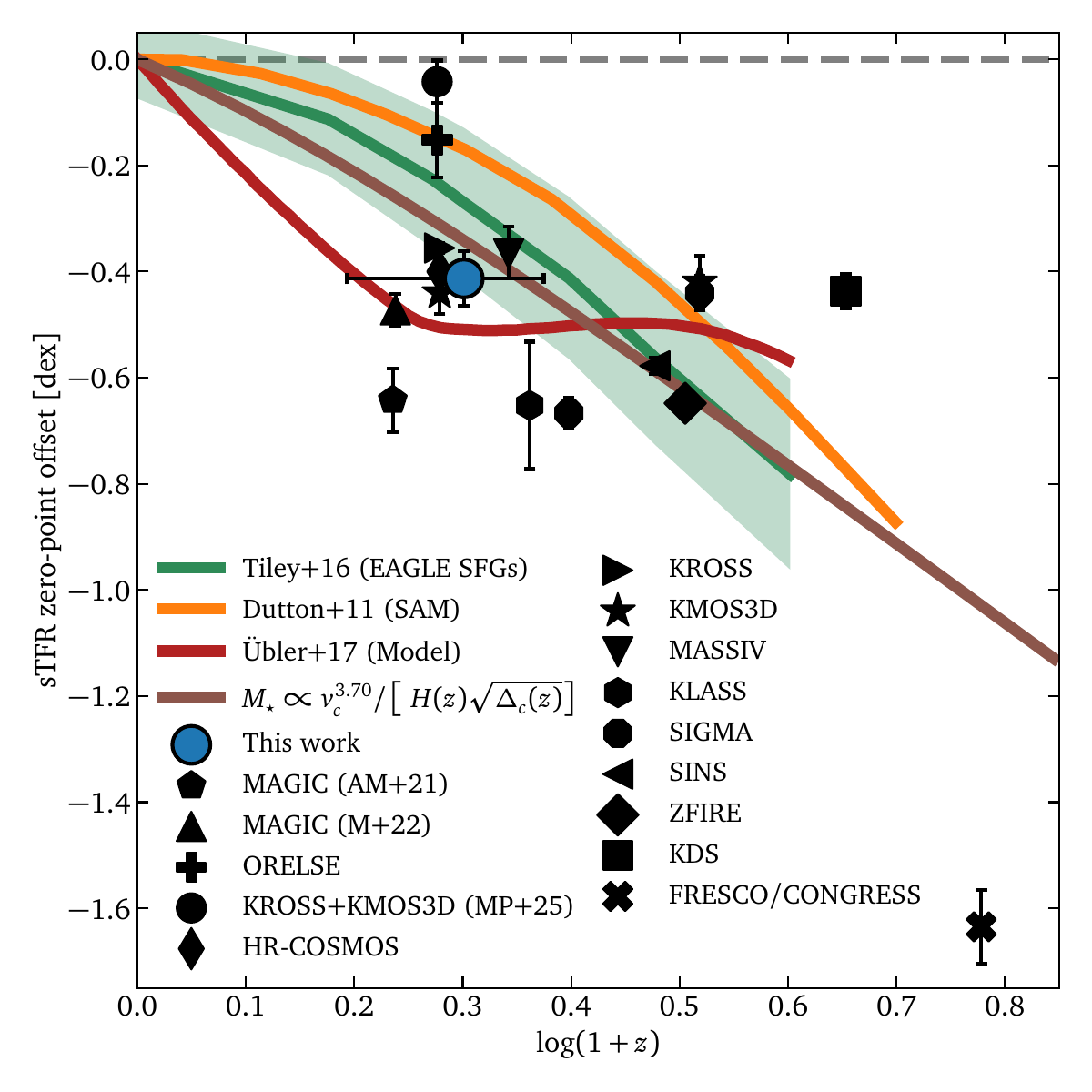}
        \caption{Evolution of the sTFR zero-point with redshift, shown relative to the local calibration of \citet{Reyes_11} along the stellar mass axis. Also plotted are predictions from the EAGLE hydrodynamical simulation (presented in \citealt{Tiley_16}, the shaded region indicates the rms dispersion of simulated SFGs), the \citet{Dutton_11} semi-analytical model, the empirical model of \citet{Ubler_17} along with their measurements on the KMOS3D sample, and the evolution attributed to cosmology. For comparison, we include a compilation of literature measurements (see text), many of which were assembled and reprocessed by \citet{Turner_17}. Horizontal error bars indicate the redshift range of our sample while vertical error bars show the fitting uncertainties for each sample.}
        \label{fig:sTFR_offset}
    \end{figure}
    
    \begin{figure}[]
        \includegraphics[width=\hsize]{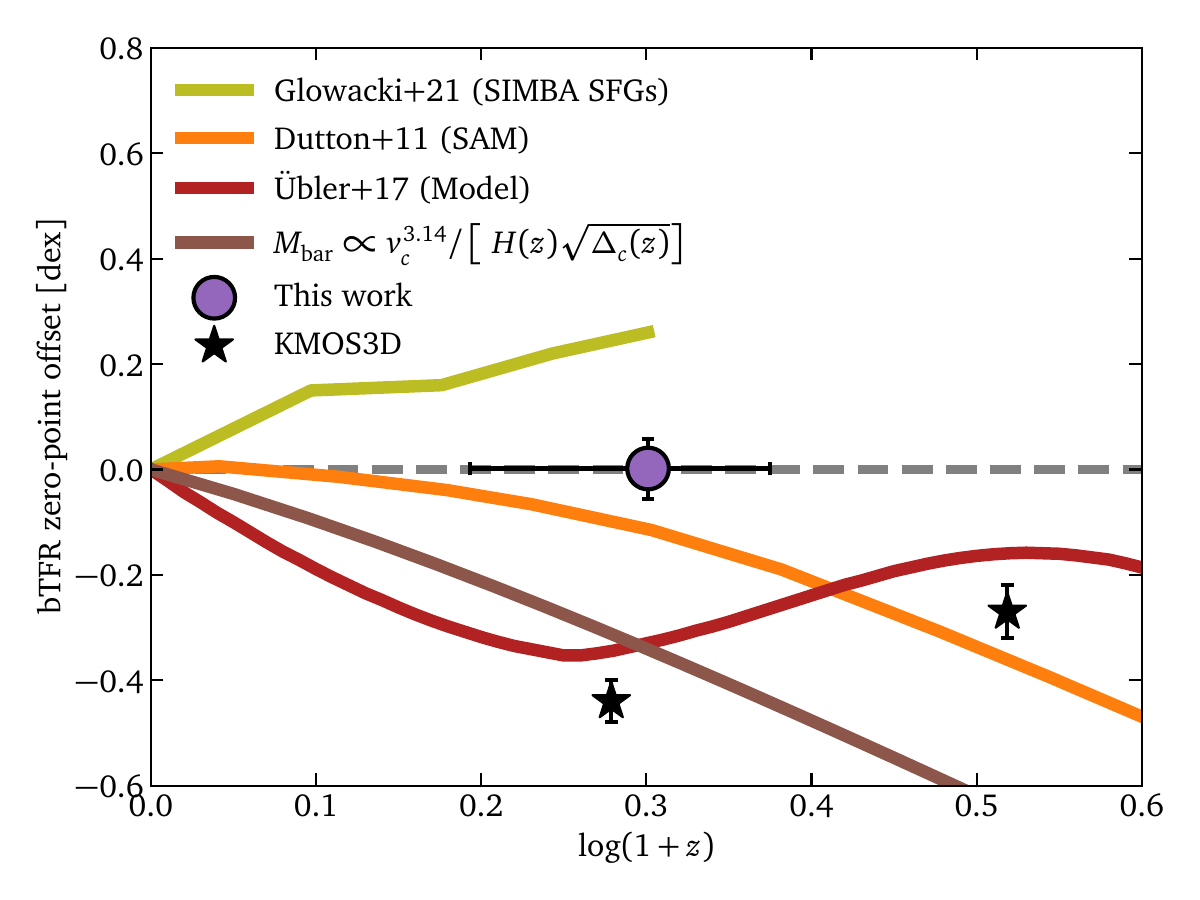}
        \caption{Evolution of the bTFR zero-point with redshift, shown relative to the local calibration of \citet{Lelli_19} along the stellar mass axis. Also plotted are predictions from the SIMBA hydrodynamical simulation \citep{Glowacki_21}, the \citet{Dutton_11} semi-analytical model, the empirical model of \citet{Ubler_17} along with their measurements on the KMOS3D sample. Horizontal error bars indicate the redshift range of our sample while vertical error bars show the fitting uncertainties for each sample.}
        \label{fig:bTFR_offset}
    \end{figure}
    
    \subsection{The TFR at $z \sim 1$}
    
        We now turn to the sTFR and bTFR of our sample, shown in Fig.~\ref{fig:TFR} together with the fiducial best-fitting relations and their posterior distributions. The corresponding parameters are listed in Tab.~\ref{tab:TFR_fits}. When the slope is allowed to vary freely, the recovered slopes for both relations are consistent with local reference values. For the sTFR, we obtain $a^\mathrm{sTFR} = 3.73^{+0.29}_{-0.26}$, consistent with the range $a^\mathrm{sTFR} = 3.6 - 3.9$ commonly reported in local studies \citep{Reyes_11, Lapi_18, Ristea_24} and in most high-redshift analyses \citep[][and references therein]{Sharma_24}. A notable exception is the significantly steeper slope ($a^\mathrm{sTFR} = 5.21 \pm 0.18$) found by \citet{Marasco_25} for a nearby sample spanning $M_\star = 10^{8-11}M_\odot$. As discussed in Appendix~\ref{appendix:ManceraPina25comparison}, their sTFR bends around $M_\star \sim 10^{10}M_\odot$ and steepens toward lower masses. For the bTFR, we measure $a^\mathrm{bTFR} = 3.11^{+0.45}_{-0.36}$, consistent with the $3.14 \pm 0.08$ slope from \citet{Lelli_19}.
        
        Given this overall consistency with local slopes, we adopt these reference values in our fiducial analysis and fit only for the zero-point and intrinsic scatter. For the sTFR, we fix the slope to the \citet{Reyes_11} value because (i) it is calibrated over $M_\star = 10^{9-11}M_\odot$, matching the projected mass range of our sample by $z \approx 0$, and (ii) it agrees with other determinations over a similar range \citep{Lapi_18, Ristea_24} as well as with our own data. This yields a zero-point $b^\mathrm{sTFR} = 1.81^{+0.05}_{-0.05}$, with an intrinsic orthogonal scatter of $\sigma^\mathrm{sTFR}_{\perp,\ \mathrm{int}} = 0.12~\mathrm{dex}$. For the bTFR, adopting the \citet{Lelli_19} slope results in $b^\mathrm{bTFR} = 3.54^{+0.06}_{-0.06}$ and $\sigma^\mathrm{bTFR}_{\perp,\ \mathrm{int}} = 0.16~\mathrm{dex}$. Relative to the corresponding local relations, the inferred zero-point offsets are $\Delta b^\mathrm{sTFR} = -0.42^{+0.05}_{-0.05}~\mathrm{dex}$ and $\Delta b^\mathrm{bTFR} = 0.00^{+0.06}_{-0.06}~\mathrm{dex}$ along the stellar mass axis. We now use our results to examine the evolution of the sTFR and bTFR zero-points in the remainder of this section. We remind that our bTFR relies on cold gas scaling relations applied to the sTFR and is therefore more indirect and subject to larger uncertainties, particularly at the low-mass end.

    \subsection{Evolution of the sTFR}
    
        In Fig.~\ref{fig:sTFR_offset}, we compare our sTFR results to intermediate and high-redshift measurements of rotation-dominated galaxies, many of which were compiled and reprocessed by \citet{Turner_17}, adopting a slope fixed to that of \citet{Reyes_11}. We use the total velocities $v_\mathrm{tot} = \sqrt{v_{\perp}^2 + 4\sigma_\mathrm{int}^2}$ from their compilation, where the rotation velocity $v_{\perp}$ is extracted at $1.3-2 R_e$ depending on the sample, and $\sigma_\mathrm{int}$ is the beam-smearing corrected velocity dispersion. This definition includes a pressure-support correction similar to that of \citet{Dalcanton_10} and is therefore directly comparable to the circular velocities used in this work, $v_{1.8} \approx \sqrt{v_{\perp}(1.8R_e)^2 + 2.8 \sigma_0^2}$. For a representative galaxy with $M_\star = 10^9 M_\odot$, $v_{\perp}(1.8R_e) = 70~\mathrm{km/s}$ and $\sigma_\mathrm{int} \approx \sigma_{0} = 30~\mathrm{km/s}$, we find $\log (v_\mathrm{tot}) - \log (v_{1.8}) = +0.03~\mathrm{dex}$, corresponding to a $\simeq -0.1~\mathrm{dex}$ shift along the stellar-mass axis in the sTFR.
        
        The comparison samples include \citet[][HR-COSMOS, $z \sim 0.9$]{Pelliccia_17}, \citet[][KROSS, $z \sim 0.9$]{Harrison_17}, \citet[][MASSIV, $z \sim 1.2$]{Epinat_12}, \citet[][SIGMA, $z \sim 1.5$ and $z \sim 2.3$]{Simons_16}, \citet[][SINS, $z \sim 2.0$]{Cresci_09}, \citet[][ZFIRE, $z \sim 2.2$]{Straatman_17} and \citet[][KDS, $z \sim 3.5$]{Turner_17}. We exclude the DYNAMO \citep{Green_14}, MKS \citep{Swinbank_17}, and AMAZE \citep{Gnerucci_11} samples, which are not directly comparable to our analysis. As discussed by \citet{Turner_17}, the DYNAMO sample is selected for high specific SFRs and gas fractions; the MKS velocities are not corrected for beam smearing, which may lead to underestimated velocities at fixed stellar mass; and the AMAZE sample, as reprocessed by \citet{Turner_17}, contains only five galaxies.
        
        We additionally include the offsets reported by \citet[][KMOS3D, $z \sim 0.9$ and $z \sim 2.3$]{Ubler_17}, whose velocities are measured at maximum ($\approx 2.2R_d$), and those reported by \citet[][FRESCO/CONGRESS, $z \sim 5$]{Danhaive_25, Danhaive_26}, with velocities measured at $R_e$. We also fitted and derived sTFR offsets for the \citet[][MAGIC, $z \sim 0.7$]{AbrilMelgarejo_21}, \citet[][MAGIC, $z \sim 0.7$]{Mercier_22} and \citet[][ORELSE, $z \sim 0.9$]{Pelliccia_19} samples, which probe galaxies across environments of varying density and measure velocities at $2.2R_d$. Finally, we fitted the sTFR offset of the \citet[][KLASS, $z \sim 1.3$]{Girard_20} and \citet[][KROSS+KMOS$^{\rm 3D}$, $z \sim 0.9$]{ManceraPina_26} samples, whose velocities are extracted in the flat part of their RCs ($\gtrsim 2R_e$). For comparison, we also show predictions from the EAGLE hydrodynamical simulation \citep[presented in][]{Tiley_16}, the semi-analytical model of \citet{Dutton_11}, the empirical model presented in \citet{Ubler_17}, and the expected evolution due to the redshift dependence of $\left[H(z)\sqrt{\Delta_c(z)}\right]^{-1}$, where $\Delta_c(z)$ is the DM halo overdensity following \citet{Bryan_98}.
        
        Our measurements are qualitatively consistent with predicted evolutions, indicating lower stellar masses at fixed velocity compared to the local relation. Quantitatively, the observed offset is most compatible with the empirical model of \citet{Ubler_17}, the EAGLE SFGs, and the expected evolution driven by the redshift dependence of halo-defining quantities. Our results are in agreement with other $z \sim 1$ studies targeting a similar mass range \citep{Mercier_22} or more massive galaxies \citep[$M_\star \sim 10^{10} M_\odot$;][]{Ubler_17, Pelliccia_17, Harrison_17, Epinat_12}, once pressure support is taken into account. This is interesting considering the various systematic differences between those samples in terms of selection (e.g. definition of rotational support, probed stellar mass range), velocity measurement (e.g. velocity definition, 2D/3D modelling, pressure-support correction) and stellar mass measurement (e.g. stellar population assumptions). While we do not engage in an extensive comparison of these systematics, we stress that our analysis relies on 3D forward modelling of the full datacubes, whereas most comparison studies model 2D velocity fields \citep{Ubler_17, Harrison_17, Epinat_12} or 2D slit spectra \citep{Pelliccia_17, Pelliccia_19}. We also note differences in the adopted criteria for rotational support, with \citet{Ubler_17} imposing a much stricter threshold of $v_{\perp,\, \mathrm{max}} / \sigma_0 > \sqrt{4.4}$, compared to $v_{\perp} / \sigma_0 > 1$ in other samples reprocessed by \citet{Turner_17} and the criterion $v_{1.8} / \sigma_0 > 1$ adopted here. We find that our conclusions are relatively robust to these choices: adopting $v_{2.0}$ and the local reference of \citet{Ristea_24} reduces the inferred evolution to $-0.37^{+0.05}_{-0.05}~\mathrm{dex}$, while imposing a stricter rotational-support criterion ($v_{1.8} / \sigma_0 > 2$) increases it to $-0.48^{+0.05}_{-0.05}~\mathrm{dex}$, at $\sim 1\sigma$ of the fiducial offset. Finally, we infer an orthogonal intrinsic scatter of $0.10 - 0.12~\mathrm{dex}$, significantly above local estimates ($0.02 - 0.08~\mathrm{dex}$). Following \citet{ManceraPina_26}, we tested whether the increased scatter could arise from underestimated velocity uncertainties. We therefore inflated the errors along the velocity axis until the fitted intrinsic scatter matched local estimates. This requires adding $\simeq 0.10~\mathrm{dex}$ in quadrature to the velocity uncertainties: an increase that would imply modelling or methodological errors \citep[e.g. the impact of non-circular motions, see][]{Dado_26} beyond the recovery performance quantified in Sect.~\ref{sect:recovery_perf} ($\simeq 0.05~\mathrm{dex}$). We also verified that inflating errors in this way does not significantly affect the inferred zero-point. Since the $0.15~\mathrm{dex}$ uncertainty on stellar mass is already intended to capture magnification and typical SED-modelling errors \citep{Pacifici_23}, and given the homogeneity of our sample, explaining the excess intrinsic scatter through additional stellar-mass errors would also imply unknown modelling or methodological uncertainties at the $\simeq 0.35~\mathrm{dex}$ level.
        
        To interpret the observed evolution of the sTFR, we relate it to the evolution of DM haloes. Assuming a spherical top-hat collapse, halo masses are defined as $M_\mathrm{vir}(z) = 4 \pi r_\mathrm{vir}^3(z) \Delta_c(z)\rho_c(z) / 3$, where $r_\mathrm{vir}(z)$ is the virial radius, within which the mean halo density is $\Delta_c(z)$ times the critical density of the Universe, $\rho_c(z) = 3 H(z)^2 / (8 \pi G)$. The evolution of the sTFR can be written \citep{Dutton_11, ManceraPina_26}
    
        \begin{equation}
        \label{eq:sTFR_evolution}
            \begin{aligned}
            \left[ \frac{M_\star(z)}{M_\star(0)} \right] = & \left[ \frac{H(z) \sqrt{\Delta_c(z)}}{H_0 \sqrt{\Delta_c(0)}}\right]^{-1} \\
            & \left[ \frac{f_M(M_\mathrm{vir}, z)}{f_M(M_\mathrm{vir}, 0)} \right] \left[ \frac{f_V(M_\mathrm{vir}, z)}{f_V(M_\mathrm{vir}, 0)} \right]^{-3} \left[ \frac{v_{1.8}(z)}{v_{1.8}(0)} \right]^3,
            \end{aligned}
        \end{equation}
    
        where $f_V = v_{1.8} / v_\mathrm{vir}$ with $v_\mathrm{vir} = v_\mathrm{vir}(M_\mathrm{vir}, z)$ the virial velocity, and $f_M = M_\star / M_\mathrm{vir}$.
    
        The first term on the right-hand side of Eq.~\ref{eq:sTFR_evolution} evolves by $-0.34~\mathrm{dex}$ between $z = 0$ and $z \sim 1$, accounting for most of the observed offset. This implies that the product $f_M f_V^{-3}$ evolves only weakly over this redshift interval. Our conclusions differ from those of \citet{ManceraPina_26}, who argue for a significant evolution through both the sTFR slope and zero-point. Given that their $z \sim 1$ slope is consistent with ours, the discrepancy in the inferred slope evolution appears to stem from their adopted local reference relation \citep{Marasco_25}, which bends and steepens toward the low-mass end. Within the overlapping mass range, however, it is broadly consistent with other $z \approx 0$ determinations \citep{Reyes_11, Lapi_18, Ristea_24} and with our results (see Appendix~\ref{appendix:ManceraPina25comparison}). The remaining difference would therefore concern the zero-point. In principle, this offset could reflect systematic shifts in stellar mass estimates driven by differences in stellar population synthesis models, dust attenuation prescriptions, or assumed SFHs. However, this explanation appears insufficient, given the homogeneous SED treatment adopted by \citet{ManceraPina_26} relative to \citet{Marasco_25}, the agreement between dynamically and photometrically inferred stellar masses reported by \citet{Marasco_25}, and, on the other hand, the body of previous studies that infer significant zero-point evolution. A severe systematic offset in velocity also seems unlikely, pointing instead to sample selection effects or other unaccounted methodological biases.
        
        We now consider whether the predicted or observed evolutions of $f_M$ and $f_V$ are quantitatively compatible with our results. The ratio $f_V$ is expected to increase modestly with cosmic time \citep{Dutton_10, Dutton_11}, reflecting the predicted increase in DM halo concentration by $\simeq +0.1-0.2~\mathrm{dex}$ since $z \sim 1$ \citep{Dutton_14, Anbajagane_22, Sorini_25} for galaxies with $M_\star \sim 10^9 M_\odot$. While this evolution remains poorly constrained observationally, initial tests of the concentration - halo mass relation at $z \sim 1$ are broadly consistent with predictions \citep{Bouche_22, Sharma_22, Ciocan_26}. Neglecting baryonic effects and assuming an NFW halo \citep{Navarro_97}, this implies an evolution of $f_V^{-3}$ of $\simeq -0.1~\mathrm{dex}$ since $z \sim 1$ \citep[][their Eq.~8]{Dutton_10}. On the other hand, abundance-matching models \citep[e.g.][]{Behroozi_19, Girelli_20} predict an increase in the stellar mass fraction $f_M$ of $\simeq +0.3~\mathrm{dex}$ since $z \sim 1$. Recent observational results from the COSMOS-Web survey, based on abundance matching \citep{Shuntov_25} or halo occupation distribution \citep{Paquereau_25} approaches, tend to show an even weaker evolution (or even hints at a larger $f_M$ at $z \sim 1$). Taken together, those expectations are consistent with a nearly constant or weakly evolving $f_M f_V^{-3}$, and an sTFR evolution dominated by the redshift dependence of halo-defining quantities since $z \sim 1$.
        
        Fig.~\ref{fig:sTFR_offset} suggests that this conclusion is supported by other samples around cosmic noon \citep{Simons_16, Cresci_09, Straatman_17, Turner_17}. ALMA and JWST now offer the opportunity to probe this picture at much earlier cosmic times. Dynamically cold discs are now identified at $z \gtrsim 4$ using ALMA observations of the \hbox{[{\rm C}{\sc \,ii}]} $\lambda$\qty{158}{\micro\metre} line \citep[e.g.][]{Rizzo_20, Rizzo_21, Fraternali_21, Herrera-Camus_22} and, using JWST/NIRCam grism spectroscopy targeting H$\alpha$ emission, \citet{Danhaive_25} report that roughly 40\% of their sample of SFGs appear rotationally supported at $z \sim 5$. Using the same sample, \citet{Danhaive_26} further suggest that the sTFR may already be emerging at that epoch. They report a comparable intrinsic scatter ($0.49~\mathrm{dex}$ in stellar mass), and measure an offset of $-1.6~\mathrm{dex}$ at $z \sim 5$ for galaxies with $M_\star > 10^8 M_\odot$, of which $-1.0~\mathrm{dex}$ can be explained by the redshift dependence of $\left[H(z)\sqrt{\Delta_c(z)}\right]^{-1}$.

    \subsection{Evolution of the bTFR}
        In Fig.~\ref{fig:bTFR_offset}, we compare our bTFR measurements with predictions from the SIMBA hydrodynamical simulation \citep{Glowacki_21}, the semi-analytical model of \citet{Dutton_11}, and the empirical model of \citet{Ubler_17}, including the offsets they report at $z \sim 0.9$ and $z \sim 2.3$ relative to the local relation of \citet{Lelli_16a}. We also show the expected evolution driven solely by the redshift dependence of $\left[H(z)\sqrt{\Delta_c(z)}\right]^{-1}$, following the same interpretation developed for the sTFR around Eq.~\ref{eq:sTFR_evolution}.

        Taking into account both the statistical and systematic uncertainties of our sample, as well as the $\pm 0.16~\mathrm{dex}$ statistical uncertainty in the local bTFR zero-point \citep{Lelli_19}, we find that our results are consistent with the predictions of the \citet{Dutton_11} model. When considered together with our sTFR results (where the observed evolution is relatively more pronounced) our findings may indicate a stronger evolution of the cold gas fraction since $z \sim 1$ relative to their model. In contrast to the SIMBA predictions of \citet{Glowacki_21}, we do not observe an increase in baryonic mass at fixed circular velocity. We note, however, that in their analysis this trend appears only when adopting velocities measured in the flat part of the RCs, and is not seen for other velocity metrics based on linewidths. We further note that their simulated sample includes low-mass galaxies ($M_\mathrm{bar} < 10^{10} M_\odot$) that deviate from the bTFR towards lower baryonic masses at fixed velocity (see their Fig.~6), a behaviour that is not present in our sample and may indicate systematic differences related to sample selection. Finally, our results differ from those of \citet{Ubler_17}, who report a moderate evolution of the bTFR since $z \sim 1$, despite the good agreement between the two studies regarding the sTFR evolution. We attribute this discrepancy to the non-negligible contribution of atomic gas to the overall baryonic mass budget in our galaxies (rather than to the mass budget within the velocity measurement radius as assumed in \citealt{Ubler_17}) based on recent empirical constraints \citep{Chowdhury_22}. Notably, both studies rely on similar scaling relations to estimate molecular gas masses.

        Investigating the scatter of the bTFR is particularly informative, as the remarkable tightness of this relation in the local Universe suggests that it may be more fundamental than other dynamical scaling relations, such as the sTFR or the Fall relations \citep{McGaugh_12, Lelli_16a, Lelli_19, Hua_25}. We find that the intrinsic orthogonal scatter of our bTFR is approximately three times larger (in dex) than that measured locally by \citet[][$0.05~\mathrm{dex}$]{Lelli_19}. Disentangling the origin of this increased scatter is, however, challenging. In addition to the larger measurement uncertainties and the enhanced contribution of non-circular motions expected at high redshift \citep[especially when using ionised gas tracers, see][]{Rizzo_24, Kohandel_24}, this likely reflects the limited precision of our cold gas mass estimates, which rely on scaling relations and constitute a limitation of our analysis. As for the sTFR, we note that one would need to inflate the velocity uncertainties by $0.15~\mathrm{dex}$ in quadrature to recover the intrinsic scatter measured locally, and that doing so does not significantly affect the inferred bTFR zero-point.

        Observational constraints on the bTFR from distant galaxies remain scarce, particularly those based on direct measurements of cold gas masses and kinematics. We highlight the first HI-based studies by \citet{Ponomareva_21} ($0 \leq z \leq 0.081$), \citet{Gogate_23} ($z \sim 0.2$), and \citet{Jarvis_25} ($z \sim 0.35$), as well as the CO-based bTFR study of \citet{Topal_18} ($0.05 \leq z \leq 0.3$), all of which report no significant evolution of the bTFR over their respective redshift ranges. \citet{Gogate_23} further examine environmental effects and find no measurable dependence of the bTFR on cluster environment. Studies of the sTFR based on ionised gas tracers similarly report little to no environmental dependence at higher redshift, provided that the samples being compared are defined and analysed homogeneously \citep[][the latter contrasting the significant offsets reported by \citealt{AbrilMelgarejo_21} and attributing them to methodological biases]{Pelliccia_19, Mercier_22}. Upcoming measurements of HI masses and linewidths from the Square Kilometre Array (SKA) and its pathfinders \citep{Staveley-Smith_15} will be key in extending bTFR constraints to earlier cosmic times.

\section{Conclusions}
\label{sect:conclusions}

    In this work, we presented an extension of the 3D forward-modelling tool \galpak{}, developed to recover the intrinsic morphology and kinematics of strongly lensed galaxies. We validated this extension using mock lensed datacubes and demonstrated its performance relative to direct image-plane fitting. We then applied this methodology to a sample of \ngal{} star-forming main-sequence galaxies at $z \sim 1$, drawn from the MUSE Atlas of Lensing Clusters and magnified by Hubble Frontier Fields clusters. For each galaxy, we forward-modelled the \OII{} emission to recover its kinematics, derived stellar population properties through SED modelling, and reconstructed the intrinsic morphology in the source plane.

    Using mock lensed datacubes representative of our sample, we showed that the intrinsic morpho-kinematic properties of strongly lensed SFGs can be reliably recovered provided they are rotation dominated ($v_{1.8}/\sigma_0 > 1$) and satisfy $S/N_\mathrm{max} \times (\sqrt{\mu} R_e / R_\mathrm{PSF})^{\alpha} \gtrsim 10$, with $\alpha \simeq 1$. We note that the recovered parameters exhibit relative uncertainties comparable to those obtained in the unlensed case at similar $\mathrm{S/N}_\mathrm{eff}$. We find that this approach is significantly more robust, particularly for morphological parameters and velocity dispersion, with improvements by factors $\simeq 2-4$ in relative errors, than methods that neglect differential magnification, even for moderate magnifications ($\mu < 6$).
    
    Restricting our analysis to \ngalfit{} galaxies that are rotationally supported ($v_{1.8} / \sigma_0 > 1$) and have well-constrained velocities ($\Delta v_{1.8} / v_{1.8} < 30\%$), we find no evidence for an evolution in the slopes of either the sTFR or bTFR. Adopting fixed slopes from local reference relations for the sTFR \citep{Reyes_11} and for the bTFR \citep{Lelli_19}, we measure a significant evolution of the sTFR zero-point, $\Delta b^\mathrm{sTFR} = -0.42^{+0.05}_{-0.05}~\mathrm{dex}$ in stellar mass, while the bTFR zero-point remains consistent with its local value, $\Delta b^\mathrm{bTFR} = 0.00^{+0.06}_{-0.06}~\mathrm{dex}$. In both cases, we find an intrinsic scatter that is larger than in the local Universe, although it remains difficult to determine whether this reflects a genuine physical evolution or unaccounted observational uncertainties.
    
    For the sTFR, our measurements are in good agreement with other $z \sim 1$ studies \citep[][]{Ubler_17, Pelliccia_17, Harrison_17, Epinat_12, Mercier_22}, once pressure support is taken into account. We tested the robustness of our results by varying the velocity definition and the adopted reference relation on the one hand, and the criterion to identify rotationally-supported galaxies on the other hand. These choices affected the inferred zero-point offset at the $\sim 1\sigma$ level but do not alter our conclusions. We also find consistency with the empirical model of \citet{Ubler_17}, predictions from EAGLE simulated SFGs, and the evolution expected from the redshift dependence of halo-defining quantities.

    Interpreting the evolution of the sTFR in the framework of DM halo growth, we conclude that the observed offset is largely driven by the evolution of the factor $\left[H(z)\sqrt{\Delta_c(z)}\right]^{-1}$. This implies that the product $f_M f_V^{-3}$, where $f_M = M_\star / M_\mathrm{vir}$ and $f_V = v_c / v_\mathrm{vir}$, has evolved only weakly since $z \sim 1$. This behaviour is consistent with current predictions for the evolution of the stellar-to-halo mass ratio and halo concentrations over the past $\sim 8~\mathrm{Gyr}$. We highlight, however, a contrasting interpretation by \citet{ManceraPina_26}, who infer a significant evolution of $f_M f_V^{-3}$ through both the sTFR slope and zero-point. We attribute the discrepancy in slope evolution to differences in the adopted local reference relations, considering that our best-fit sTFR slope at $z \sim 1$ ($a^\mathrm{sTFR} = 3.75^{+0.30}_{-0.27}$) is compatible with theirs ($a^\mathrm{sTFR} = 3.82^{+0.55}_{-0.40}$). The remaining zero-point offset is difficult to attribute to systematic shifts in stellar masses or circular velocities arising from measurement or modelling choices, and may instead point to residual sample selection or methodological biases.

    Finally, the absence of detectable evolution in the bTFR zero-point suggests that the increasing contribution of cold gas mass at higher redshift fully compensates the evolution observed in the stellar component alone. The increased intrinsic scatter of the bTFR probably highlights the current limitations imposed by indirect cold gas mass estimates based on scaling relations. Future high-redshift HI surveys will be valuable to further constrain the physical origin and cosmic evolution of both the sTFR and bTFR, providing direct measurements of the cold gas reservoir as well as homogeneous kinematics measurements that probe the outermost regions.

\begin{acknowledgements}
      This work utilises gravitational lensing models produced by PIs Natarajan \& Kneib (CATS), Sharon, Keeton, Diego, and the GLAFIC group. This lens modeling was partially funded by the HST Frontier Fields program conducted by STScI. STScI is operated by the Association of Universities for Research in Astronomy, Inc. under NASA contract NAS 5-26555. The lens models were obtained from the Mikulski Archive for Space Telescopes (MAST). Beyond those already cited, this work makes use of the following open-source Python libraries: \texttt{Astropy} \citep{Astropy_22}, \texttt{NumPy} \citep{Harris_20}, \texttt{matplotlib} \citep{Hunter_07}, \texttt{MPDAF} \citep{Piqueras_17}, \texttt{SciPy} \citep{Virtanen_20} and \texttt{corner} \citep{Foreman-Mackey_16}. NB and BC acknowledge support from the ANR DARK grant (ANR-22-CE31-0006).
\end{acknowledgements}

\bibliographystyle{aa_url}
\bibliography{references}

\begin{appendix}

\section{Summary of the MUSE observations and ancillary datasets}
\label{appendix:clusters}
    Tab.~\ref{tab:clusters} summarises the MUSE observations and ancillary datasets used in this work (Sect.~\ref{sect:data}). Note that the MACS0416 North (N) and South (S) pointings have been analysed separately.

    \begin{table*}[]
    \centering
    \caption{Summary of the MUSE observations and ancillary datasets.}
    \label{tab:clusters} 
    \begin{tabularx}{\textwidth}{lcccCcc}
        \hline
        \hline
        \noalign{\smallskip}
        Cluster & Pointings & $\mathrm{T_{eff}}$ & PSF FWHM & Photometry filters & Ref.MUSE & Ref.Lens \\
        &  & (hr) & (arcsec) & & &\\
        \noalign{\smallskip}
        \hline
        \noalign{\smallskip}
        \noalign{\smallskip}
        A370 & 4 & 1.5-8.5 & 0.66 & F275W, F336W, F435W, F475W, F606W, F625W, F814W, F105W, F110W, F125W, F140W, F160W, K$_{\rm S}$, \qty{3.6}{\micro\metre}, \qty{4.5}{\micro\metre}, \qty{5.8}{\micro\metre}, \qty{8.0}{\micro\metre} & R21 & L19 \\
         \noalign{\smallskip}
         A2744 & 4 & 3.5-7 & 0.61 & F275W, F336W, F435W, F606W, F814W, F105W, F125W, F140W, F160W, K$_{\rm S}$, \qty{3.6}{\micro\metre}, \qty{4.5}{\micro\metre}, \qty{5.8}{\micro\metre}, \qty{8.0}{\micro\metre} & R21 & M18\\
         \noalign{\smallskip}
         MACS0416 & 1 (N) + 1 (S) & 17 (N), 11-15 (S) & 0.53 (N), 0.65 (S) & F225W, F275W, F336W, F390W, F435W, F475W, F606W, F625W, F775W, F814W, F850LP, F105W, F110W, F125W, F140W, F160W, K$_{\rm S}$, \qty{3.6}{\micro\metre}, \qty{4.5}{\micro\metre} & R21 & R21\\
         \noalign{\smallskip}
         AS1063 & 2 & 3.9 & 1.02 & F225W, F275W, F336W, F390W, F435W, F475W, F606W, F625W, F775W, F814W, F850LP, F105W, F110W, F125W, F140W, F160W, K$_{\rm S}$, \qty{3.6}{\micro\metre}, \qty{4.5}{\micro\metre}, \qty{5.8}{\micro\metre}, \qty{8.0}{\micro\metre} & C22 & B24\\
         \noalign{\smallskip}
         \hline
         \noalign{\smallskip}\\
    \end{tabularx}
    \tablefoot{
    Dataset;
    Number of MUSE pointings;
    Range of effective MUSE exposure time per spaxel;
    PSF FWHM of the combined MUSE datacube at \qty{700}{\nano\metre};
    HST/WFC3, HST/ACS, VLT/HAWK-I and Spitzer/IRAC filters included in the HFF-DeepSpace photometric catalogue;
    References for the MUSE data;
    References for the lens model.
    \\
    L19: \citet{Lagattuta_19}\\
    M18: \citet{Mahler_18}\\
    R21: \citet{Richard_21}\\
    C22: \citet{Claeyssens_22}\\
    B24: \citet{Beauchesne_24}
    } 
    \end{table*}

\section{Properties of the mock sample}
\label{appendix:mock_properties}
\begin{figure*}[]
        \includegraphics[width=\textwidth]{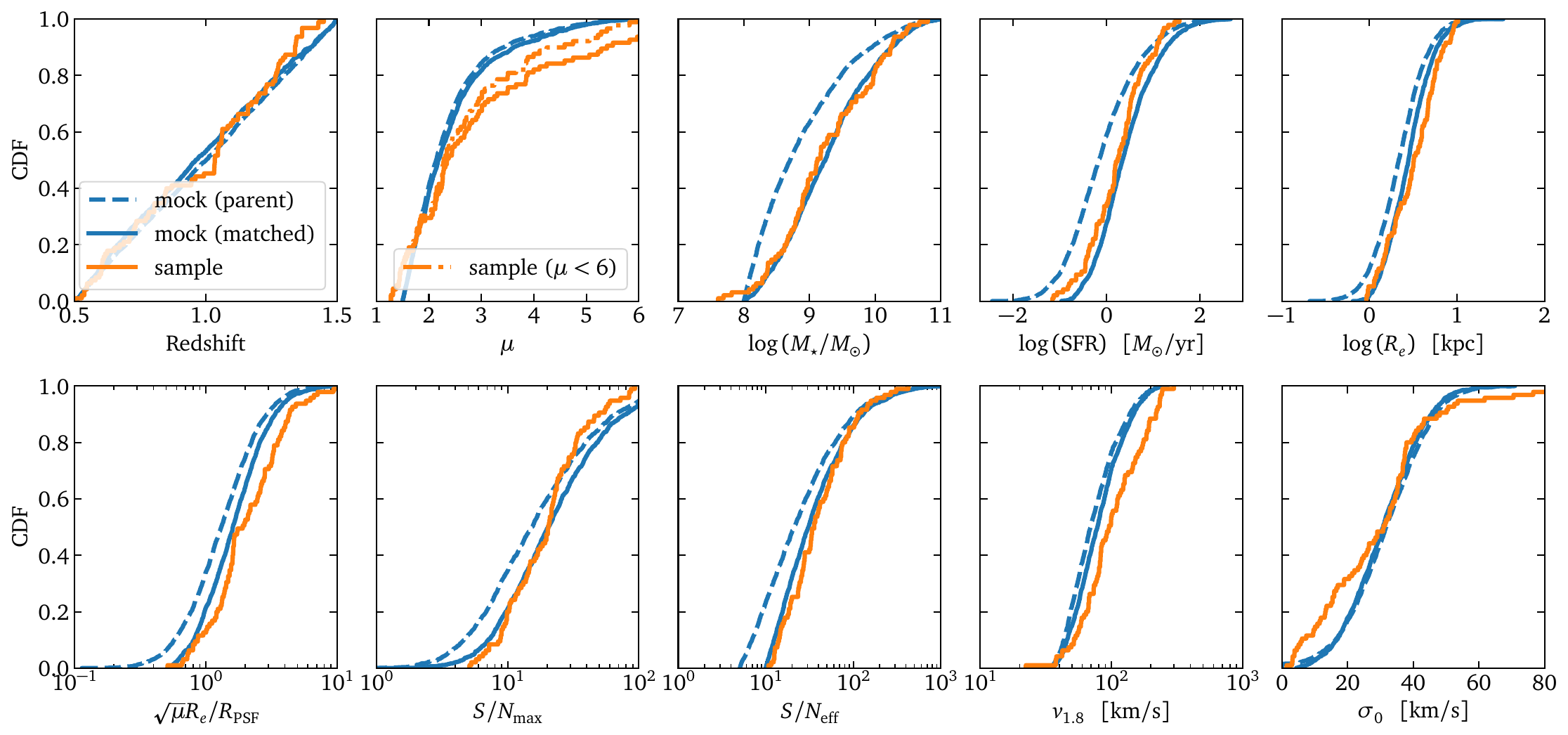}
        \caption{Properties of the mock sample compared to the kinematic sample. From top left to bottom right, we show the cumulative distribution functions (CDFs) of redshift, magnification, stellar mass, SFR, effective radius, apparent size-to-PSF ratio, S/N in the brightest spaxel, effective S/N, circular velocity at $1.8 R_e$ and velocity dispersion floor. The rotation-dominated ($v_{1.8}/\sigma_0 > 1$) subset of the kinematic sample with well-constrained velocities ($\Delta v_{1.8}/v_{1.8} < 30\%$) is shown in solid orange. A further restriction to $\mu < 6$, applied to match the magnification range of the mock sample for a fair comparison of the magnification CDFs, is indicated by the dash-dotted orange line. The parent mock sample appears as a dashed blue line, while the mock subsample satisfying selection criteria similar to the kinematic sample ($\mathrm{S/N}_\mathrm{eff} > 10$, $\sqrt{\mu} R_e /R_\mathrm{PSF} > 1/2$ and $v_{1.8} / \sigma_0 > 1$) is shown in solid blue.}
        \label{fig:mock_properties}
    \end{figure*}

    Fig.~\ref{fig:mock_properties} compares the properties of the mock sample (Sect.~\ref{sect:recovery_perf}) with those of the kinematic sample (Sect.~\ref{sect:sample}). After applying similar selection cuts, the distributions of most properties in the mocks closely match the kinematic sample, except the highest circular velocities and lowest velocity dispersions, which are both under-represented in the mock sample. 
    
\section{The impact of cluster mass modelling errors}
\label{appendix:mass_modelling}
    
    The \galpak{} Strong Lensing Extension described in Sect.~\ref{sect:galpak_SLextension} uses deflection maps derived from a lens model to account for the lensing distortions. Therefore, the inferred morpho-kinematic properties are affected by lens model errors over which our extension does not marginalise. In this Appendix, we assess the impact of those errors by comparing the circular velocities obtained with the fiducial model from \citetalias{Richard_21} to those derived using an average model constructed from five publicly available models\footnote{\url{https://archive.stsci.edu/prepds/frontier/lensmodels/}}: the \citetalias{Richard_21} model together with models produced by PIs Sharon, Keeton, Diego, and the GLAFIC group (as described in \citealt{Lotz_17}). Instead of directly averaging magnification values (which are highly non-linear), we constructed this average model by re-gridding the available maps (deflection along both axes, potential, convergence and projected shear along both axes) to the same reference frame and averaging them. 

    \begin{figure}[]
        \centering
        \includegraphics[width=\hsize]{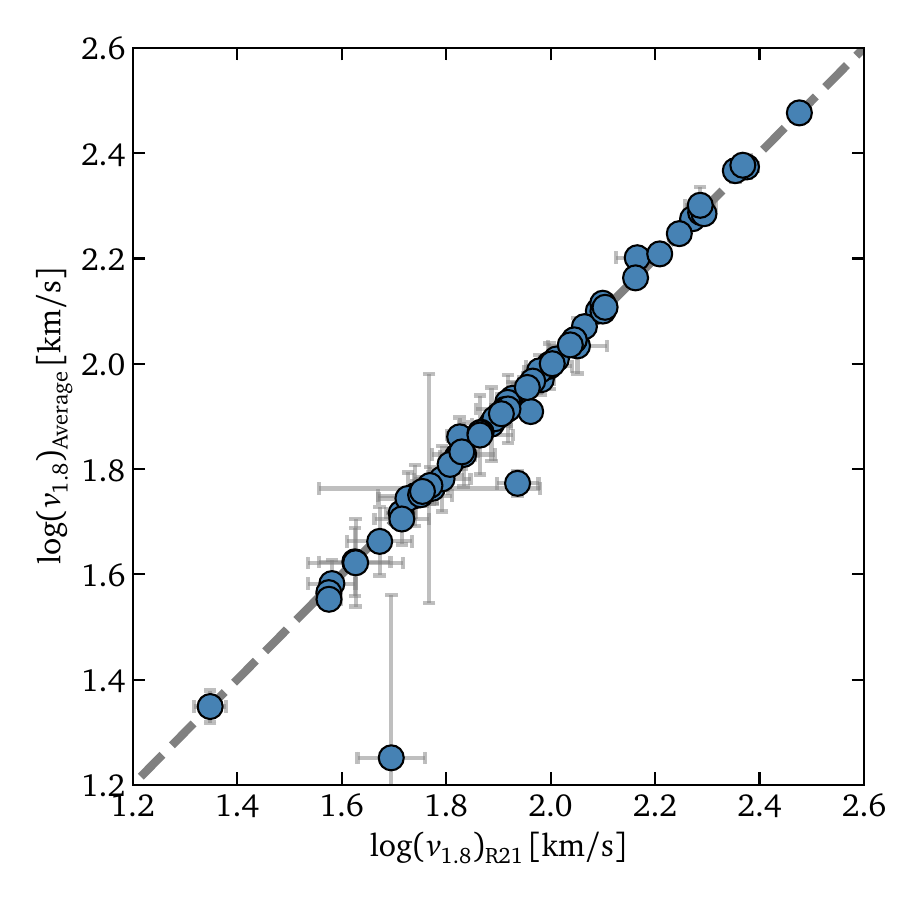}
        \caption{Comparison of circular velocities derived using the \citetalias{Richard_21} lens model adopted in this work with those obtained from an average lens model constructed from publicly available models including \citetalias{Richard_21}.}
        \label{fig:avg_v_cats_lensmodel}
    \end{figure}
    
    From our fiducial analysis, we retain \ngalfit{} galaxies that are rotationally supported ($v_{1.8} / \sigma_0 > 1$) and have well-constrained velocities ($\Delta v_{1.8} / v_{1.8} < 30\%$). We remove 24 galaxies which were included as lenses in at least one lens model, and extract the morpho-kinematics of the remaining galaxies using the same modelling assumptions and priors as for the fiducial lens model.
    
    Fig.~\ref{fig:avg_v_cats_lensmodel} compares the circular velocities obtained with the average and \citetalias{Richard_21} lens models and shows overall excellent agreement. We find a median absolute deviation of $0.003~\mathrm{dex}$, with a single outlier at $0.4~\mathrm{dex}$. This galaxy (ID4517) lies close to the critical lines of Abell~370 ($\mu = 30$–$60$), where cluster mass models are known to be particularly sensitive to modelling uncertainties. Overall, this comparison indicates that the circular velocities derived in this work are robust to uncertainties in the cluster mass modelling.

\section{Comparison with the sTFR slope evolution reported by \citet{ManceraPina_26}}
\label{appendix:ManceraPina25comparison}
\begin{figure}[]
        \centering
        \includegraphics[width=\hsize]{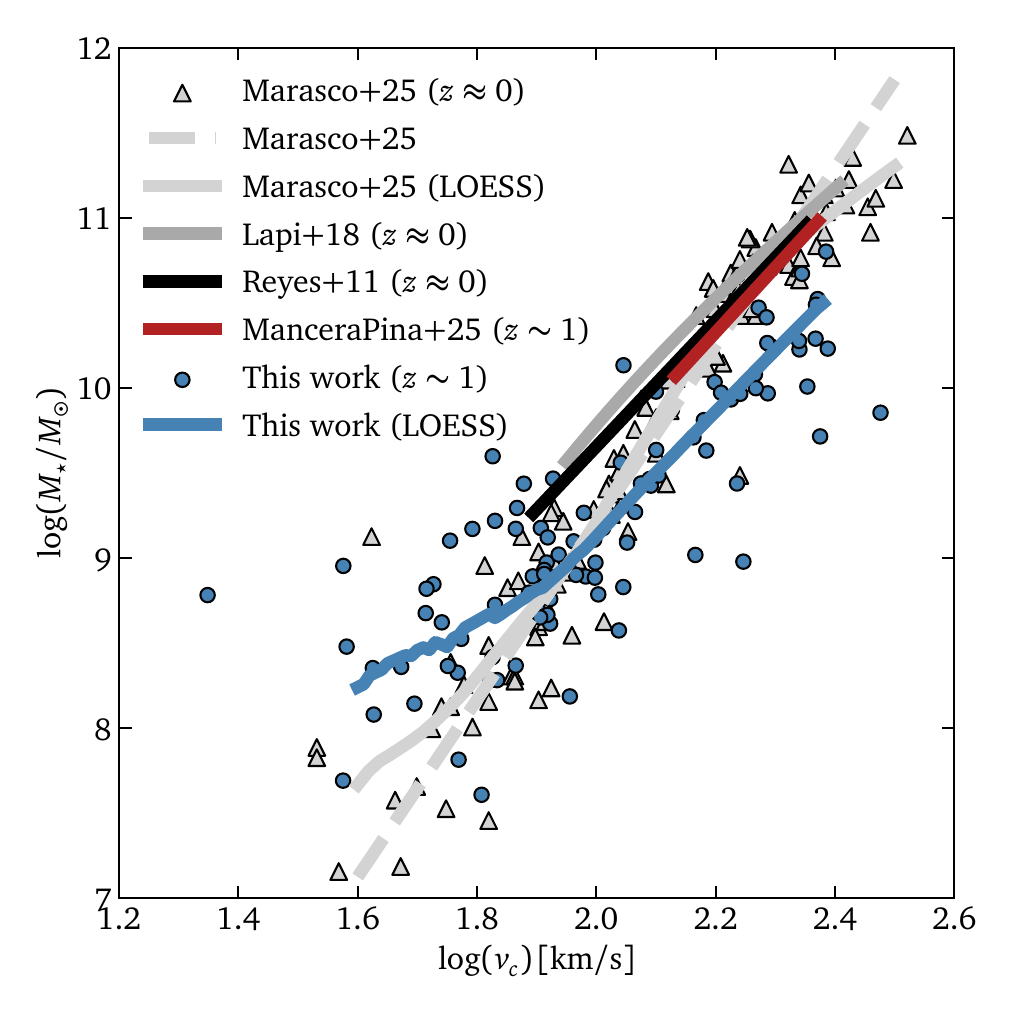}
        \caption{Comparison of the sTFR at $z \sim 1$ from \citetalias{ManceraPina_26} and this work, alongside $z \approx 0$ reference relations \citep{Reyes_11, Lapi_18, Marasco_25}. Our sample and that of \citet{Marasco_25} are shown as blue and grey symbols, respectively, with LOESS-smoothed curves \citep{Cappellari_13} highlighting the trends. The unbroken power-law sTFR from \citet{Marasco_25} is shown as a dashed grey line.}
        \label{fig:sTFR_ManceraPina25comparison}
    \end{figure}

    In this Appendix, we compare the stellar masses and circular velocities measured in this work with those reported by \citet[][hereafter MP26]{ManceraPina_26}. \citetalias{ManceraPina_26} find moderate evidence for a shallower sTFR slope at $z \sim 1$ compared to the $z \approx 0$ relation of \citet{Marasco_25}, implying a shallower $f_M - M_\star$ relation if $f_V$ is assumed independent of stellar mass and redshift. Fig.~\ref{fig:sTFR_ManceraPina25comparison} shows the sTFR from \citetalias{ManceraPina_26} and our sample at $z \sim 1$, together with $z \approx 0$ reference relations \citep[][polynomial fit]{Reyes_11, Marasco_25, Lapi_18} limited to the stellar mass range of the corresponding data. The \citet{Marasco_25} sample spans $M_\star = 10^{8-11}M_\odot$ and is consistent with previous results for $\log(v_c / [\mathrm{km~s^{-1}}]) > 2.15$. At lower velocities, the sTFR steepens, indicating lower stellar masses at fixed circular velocity. While this break has previously been reported around $\log(v_c / [\mathrm{km~s^{-1}}]) \simeq 1.95$ \citep[e.g.][]{McGaugh_00, Torres-Flores_11}, in this sample the departure occurs at higher velocities. Within the overlapping mass range, we do not observe a departure in slope between our results, \citetalias{ManceraPina_26}, and previous $z \approx 0$ determinations, but mostly a difference in zero-point.
    
\end{appendix}
\label{LastPage}
\end{document}